\documentclass[nonacm]{acmart}

\usepackage{hyperref} 
\usepackage{url}

\usepackage{listings}
\usepackage{amsmath}
\usepackage{graphicx}
\usepackage{mdframed}
\usepackage{framed}
\usepackage{wrapfig}
\usepackage{subcaption}

\usepackage{etoolbox}
\usepackage{enumitem} 

\usepackage{colortbl}

\usepackage{multirow}
\usepackage{tabularx}
\usepackage{alltt}

\usepackage{tikz}

\DeclareRobustCommand{\blackcircled}[1]{%
  {\small
  \tikz[baseline=(X.base)]{
    \node[
      circle,
      draw,
      fill=white,
      inner sep=0pt,
      outer sep=0pt,
      minimum size=0.9em,
      text width=0.9em,
      align=center,
      text height=1.5ex,
      text depth=.25ex
    ] (X) {#1};
  }}%
}
\DeclareRobustCommand{\graysquare}{%
  \raisebox{0.6ex}{%
    \begin{tikzpicture}[baseline={(current bounding box.center)}]
      \fill[gray!45] (0,0) rectangle (1.5ex,1.5ex);
    \end{tikzpicture}%
  }%
}
\DeclareRobustCommand{\graytriangle}{%
  \raisebox{0.6ex}{
    \begin{tikzpicture}[baseline={(current bounding box.center)}]
      \fill[gray!45] (0,0) -- (0.95ex,1.6ex) -- (1.9ex,0) -- cycle;
    \end{tikzpicture}
  }
}
\usetikzlibrary{}

\newcommand{\firstmode}{\textsc{{Divergent}}}
\newcommand{\secondmode}{\textsc{{Convergent}}}

\newcommand{\ours}{\textsc{HAICo}}

\definecolor{promptinputcolor}{rgb}{0,0,2.55}

\AtBeginDocument{%
  }

\setcopyright{acmlicensed}
\copyrightyear{2018}
\acmYear{2018}
\acmDOI{XXXXXXX.XXXXXXX}
\acmConference[Conference acronym 'XX]{Make sure to enter the correct
  conference title from your rights confirmation email}{June 03--05,
  2018}{Woodstock, NY}
\acmISBN{978-1-4503-XXXX-X/2018/06}

\author{Chao Wen}
\email{chaowen@mpi-sws.org}
\affiliation{%
  \institution{Max Planck Institute for Software Systems}
  \country{Germany}
}

\author{Tung Phung}
\email{mphung@mpi-sws.org}
\affiliation{%
  \institution{Max Planck Institute for Software Systems}
  \country{Germany}
}

\author{Pronita Mehrotra}
\email{pronita@mindantix.com}
\affiliation{%
  \institution{MindAntix}
  \country{USA}
}

\author{Sumit Gulwani}
\email{sumitg@microsoft.com}
\affiliation{%
  \institution{Microsoft}
  \country{USA}
}

\author{Roger E. Beaty}
\email{rebeaty@psu.edu}
\affiliation{%
  \institution{Pennsylvania State University}
  \country{USA}
}

\author{Tomohiro Nagashima}
\email{nagashima@cs.uni-saarland.de}
\affiliation{%
  \institution{Saarland Informatics Campus, Saarland University}
  \country{Germany}
}

\author{Adish Singla}
\email{adishs@mpi-sws.org}
\affiliation{%
  \institution{Max Planck Institute for Software Systems}
  \country{Germany}
}

\authorsaddresses{Preprint. Corresponding author: Chao Wen <\nolinkurl{chaowen@mpi-sws.org}>.}

\renewcommand{\shortauthors}{Wen et al.}

\begin{document}

\title[Exploration vs. Fixation: Scaffolding Divergent and Convergent Thinking for Human-AI Co-Creation]{Exploration vs. Fixation: Scaffolding Divergent and Convergent Thinking for Human-AI Co-Creation with Generative Models}

\begin{abstract}

\looseness-1
Generative AI has democratized content creation, but popular chatbot-based interfaces often prioritize execution, generating fully rendered artifacts right away. This issue can lead to premature convergence and design fixation, where users are being anchored to initial outputs. Recent works have proposed new interfaces to address this issue by supporting exploration, though typically constrained to be semantically close to a user's initial task framing, potentially limiting the creativity of the outcomes. We examine an approach grounded in the Geneplore model of creative cognition and instantiate it in a human-AI co-creation system, \ours{}, for creative image generation. \ours{} explicitly structures the creative process into two switchable modes: \firstmode{} mode scaffolds the broad exploration of remote conceptual ideas; \secondmode{} mode supports a targeted refinement of selected ideas. Through a within-subjects study ($N=24$) on a poster image creation task, we demonstrate that \ours{} outperforms ChatGPT across multiple dimensions of creativity and usability. Our results highlight the critical need to shift from pure execution-focused chatbots to scaffolded co-creation systems that actively guide exploration and foster the creative process.

\end{abstract}


\begin{teaserfigure}
    \centering
    \includegraphics[width=\textwidth]{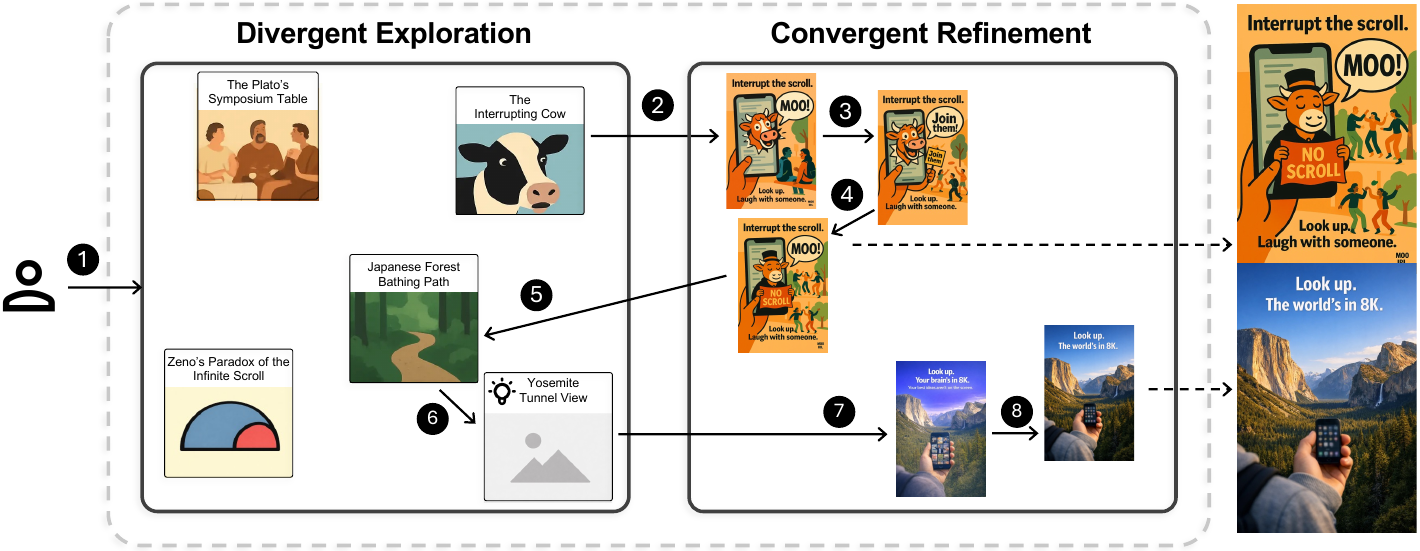}
    \caption{
        \looseness-1 Example user trajectory in our system workflow for image co-creation. 
        The numbers indicate the chronological steps taken by a user. \blackcircled{1} The user prompts the system to create a poster image about ``Spend Less Time on Phones,'' then \blackcircled{2} selects a system-initiated idea to generate a first image. After \blackcircled{3}\blackcircled{4} refining that image over two iterations, \blackcircled{5} the user returns to explore other ideas and \blackcircled{6} draws on another system-initiated idea as inspiration to develop a new idea, ``Yosemite Tunnel View.'' \blackcircled{7} The user generates an image based on their own idea and \blackcircled{8} refines it into a second result. This trajectory illustrates how users move back and forth between exploring ideas and refining promising images.
}
    \label{fig.teaser}
\end{teaserfigure}

\maketitle


\section{Introduction} \label{sec:introduction}

Generative AI has fundamentally democratized content creation, enabling novices to produce artifacts such as code, images, and videos that previously required professional expertise and complex tools~\cite{epstein2023art,huh2025videodiff,tilekbay2024expressedit,DBLP:conf/chi/TaoLPWPD25,gadde2025democratizing,DBLP:conf/cvpr/RombachBLEO22,DBLP:conf/iclr/NijkampPHTWZSX23}.
However, the widely used chatbot-based interfaces (e.g., ChatGPT) often prioritize execution, generating fully rendered artifacts immediately. 
Prior studies have found that these interfaces can lead to premature convergence and design fixation, where users are anchored to initial outputs without exploring the broader space of possibilities~\cite{DBLP:conf/chi/Wadinambiarachchi24,DBLP:conf/chi/SuhCMLX24}.
Furthermore, when refining these initial results, design fixation often persists, as users typically settle on a single refined result instead of branching into and exploring alternative variations that may lead to better outcomes. This refinement process is further exacerbated by the ``gulf of envisioning,'' where users struggle to translate their intentions into prompts that accurately instruct generative models, leading to arbitrary model behaviors and frustrating trial-and-error~\cite{DBLP:conf/chi/Zamfirescu-Pereira23,DBLP:conf/chi/SubramonyamPPAS24}.

\looseness-1
Recent systems have attempted to support exploration and refinement for content creation, but remain limited in three respects. First, systems designed to support divergent exploration mostly surface dimensions within the user's existing task framing~\cite{DBLP:conf/chi/SuhCMLX24,DBLP:conf/chi/TaoLPWPD25,DBLP:conf/chi/WangLC0C25, DBLP:conf/chi/ChoiHPCK24} or draw analogies structured around the user's design task~\cite{10.1145/3706598.3713375,DBLP:conf/chi/LinKMKCH25}. However, the exploration space is typically constrained to be semantically close to a user's initial task framing, potentially limiting the creativity of the outcomes.
Creativity research suggests that creativity often involves linking remote, semantically distant concepts~\cite{beaty2023associative,mednick1962associative}, a mechanism that neither line of work fully operationalizes.
Second, existing systems support refinement mostly by improving prompt fidelity, assuming the user already holds a concrete direction; they do not address the upstream problem of forming that direction in the first place~\cite{DBLP:conf/uist/BradeWSOG23,DBLP:conf/chi/WangHS0024,DBLP:conf/uist/WangLLHCCC025}.
Finally, many of these systems support either exploration or refinement, but still provide only limited support for the non-linear nature of creative processes, in which users move back and forth between divergent thinking for broad ideation and convergent thinking for refining a specific idea.

\looseness-1
To address these limitations, we examine an approach grounded in the Geneplore model of creative cognition~\cite{finke1996creative,PATTERSON2004843}. This approach explicitly structures interaction into two switchable modes: (i) the \firstmode{} mode scaffolds divergent thinking for generating diverse ideas and (ii) the \secondmode{} mode scaffolds convergent thinking for exploring how an idea can be adapted for solving the original task. We instantiate this approach in \ours{}, a \underline{H}uman-\underline{AI} \underline{Co}-creation system for creative image generation. \ours{} is structured around these two explicit modes.
In the \firstmode{} mode, \ours{} populates an interactive idea grid with conceptual ideas drawn from remote, cross-domain sources, supporting creative ideation outside the user's initial task framing.
In the \secondmode{} mode, \ours{} translates the user's refinement/adaptation intention into named semantic parameters with structured, context-aware dropdown options for selection, clarifying what the user wants to change and making the space of possible modifications legible.
These two modes are designed to be switchable, supporting the non-linear nature of creative work.
To investigate the effectiveness of \ours{}, we evaluate \ours{} against ChatGPT in a within-subjects study ($N=24$) on a poster image creation task. Our results demonstrate that \ours{} outperforms ChatGPT across multiple dimensions of creativity and usability. 
In summary, our contributions are as follows:
\begin{itemize}[leftmargin=*]
    \item \looseness-1We examine an approach grounded in the Geneplore model and instantiate it in \ours{}, an image co-creation system with two switchable modes: \firstmode{} mode for broad exploration by surfacing remote, cross-domain concepts, and \secondmode{} mode for structured refinement via semantic parameters and options.
   
    %
    \item \looseness-1We conduct a study with 24 participants and provide empirical evidence that \ours{} enables the creation of significantly more novel and diverse images than ChatGPT, and has higher perceived creativity support and system usability.

    %
    \item We present preliminary findings that scaffolded co-creation can foster learning through usage, with significantly higher self-reported learning in \ours{} characterized by task-specific knowledge rather than tool-specific strategies.

\end{itemize}

\section{Related Work}\label{sec.relatedwork}

We discuss prior work related to our research, focusing on cognitive research on creativity and systems that support human-AI co-creation and text-to-image generation.

\subsection{Divergent Thinking and Convergent Thinking for Creativity}

\looseness-1 Divergent and convergent thinking are widely regarded as two core cognitive processes that are central to creativity~\cite{guilford1956structure,guilford1967nature}. Divergent thinking focuses on generating a wide range of varied ideas in response to a question or task. In contrast, convergent thinking focuses on refining those high-level ideas into viable solutions~\cite{guilford1956structure}. 

\looseness-1 Previous research has established that creative models often unfold as transitions between divergent and convergent thinking~\cite{CreativeProblemSolving}. For instance, Finke et al.'s Geneplore model~\cite{finke1996creative} describes creative cognition as involving two processes: 
in the \textit{generative} process, individuals produce diverse high-level ``preinventive'' structures without strong goal control (i.e., divergent thinking).
In the \textit{exploratory} process, these structures are interpreted and developed with respect to the specific task (i.e., convergent thinking).
The \emph{generative} process may involve several cognitive mechanisms, with one central mechanism being association: the process of linking concepts stored in memory~\cite{beaty2023associative,mednick1962associative}.
The ability to form such associations is closely related to human creativity. Creativity research shows that highly creative people are more associative in their thinking: they travel further in semantic space, switch between more semantic subcategories, and make larger leaps between associations~\cite{beaty2023associative}.

\looseness-1 Inspired by these cognitive theories of creativity, we examine an approach grounded in the Geneplore model to support divergent and convergent thinking for creative tasks. We instantiate this approach in \ours{}, a system for human-AI co-creation of images. In our system, the \emph{generative} process is powered by an associative-thinking prompting strategy that draws remote connections from distant domains to generate preinventive structures~\cite{DBLP:journals/corr/abs-2405-06715}.

\subsection{Supporting Human-AI Co-Creation}

Generative AI has democratized content creation, enabling novices to produce complex artifacts such as educational exercises~\cite{chao2024xlogo,DBLP:conf/aied/GhoshTDS22}, code~\cite{DBLP:conf/acl/WenSS25,DBLP:journals/corr/abs-2308-12950}, and images~\cite{DBLP:conf/uist/BradeWSOG23,DBLP:conf/chi/WangLC0C25}. Early work primarily framed generative models as productivity tools that automate content creation~\cite{DBLP:journals/corr/abs-2107-03374,chao2024xlogo,DBLP:journals/corr/abs-2406-09961,DBLP:conf/acl/WenSS25}. More recent research has begun to frame generative models as collaborators and design AI systems for creative tasks~\cite{DBLP:conf/chi/SuhCMLX24,DBLP:conf/chi/TaoLPWPD25,10.1145/3706598.3713375,DBLP:conf/chi/WangLC0C25,DBLP:conf/chi/ChoiHPCK24}.

\looseness-1
To support human-AI co-creation, recent systems design interaction mechanisms and interfaces that facilitate exploration, often by helping users navigate a design space through structured dimensions or analogical inspiration.
Dimension-based systems such as Luminate~\cite{DBLP:conf/chi/SuhCMLX24}, DesignWeaver~\cite{DBLP:conf/chi/TaoLPWPD25}, AIdeation~\cite{DBLP:conf/chi/WangLC0C25}, and CreativeConnect~\cite{DBLP:conf/chi/ChoiHPCK24} surface task-relevant attributes such as genre, tone, art style, and visual keywords, helping users navigate alternatives within their existing task framing. Analogy-based systems such as IdeationWeb~\cite{10.1145/3706598.3713375} and Inkspire~\cite{DBLP:conf/chi/LinKMKCH25} go further by drawing inspiration from external source domains to help users move beyond their initial mental model.

\looseness-1However, as mentioned above, creative ideation benefits from remote associative stimuli sufficiently distant from the user's starting point to catalyze genuinely new directions~\cite{beaty2023associative,mednick1962associative,dahl2002influence,casakin1999expertise}. Dimension-based systems operate on attributes already adjacent to the user's prompt, while analogy-based systems draw from fixed or domain-constrained source categories targeting expert designers in specific professional settings. Neither leverages an exploration space broad enough to introduce ideas genuinely remote from the user's initial mental model and task framing. Within \ours{}, the \firstmode{} mode addresses this issue by operationalizing associative thinking through prompting generative models to surface remote conceptual ideas spanning various domains, such as mythology, historical events, and internet culture.

\subsection{Supporting Text-to-Image Generation}

Another line of work designs systems to support text-to-image generation by helping users articulate, inspect, and refine prompts to better control generative model outputs. These systems improve prompt engineering through mechanisms such as prompt expansion, keyword recommendation, prompt optimization, and visual feedback on model behavior. For example, Promptify~\cite{DBLP:conf/uist/BradeWSOG23} supports expanding user prompts with additional subjects and stylistic details; PromptCharm~\cite{DBLP:conf/chi/WangHS0024} automatically refines user prompts and supports iterative image improvement through attention visualization and inpainting; GenTune~\cite{DBLP:conf/uist/WangLLHCCC025} enables traceable prompt refinement by linking image elements back to their corresponding prompt labels for precise, globally consistent edits; and PromptMagician~\cite{DBLP:journals/tvcg/FengWWWLZWC24} supports interactive prompt refinement through keyword recommendation, prompt-image retrieval, and prompt-space visualization.
These systems are primarily designed to improve the fidelity between user intent and model output.

However, these systems are generally effective only when users enter the interaction with a concrete direction already in mind. They do not address the challenge of forming that direction in the first place. Consequently, improved prompt fidelity may improve specification of an existing idea, but it does not expand the exploration space by introducing directions the user may not have previously considered. Furthermore, even when these systems surface alternatives, they remain anchored at the \emph{prompt level}, conditioned on the vocabulary of the current prompt~\cite{DBLP:conf/uist/BradeWSOG23,DBLP:journals/tvcg/FengWWWLZWC24}. 
Our work differs in two key ways. First, \ours{} introduces a dedicated \firstmode{} mode prior to image generation, enabling divergent exploration of conceptual directions before committing to an image. Second, the \secondmode{} mode surfaces alternatives at the \emph{intent level} (i.e., what users want to achieve) rather than at the prompt level.


\section{Problem Analysis and Design Goals}
\label{sec:problemsetup}

For human-AI co-creation, current interaction paradigms, especially for novices, typically follow a workflow where the user submits a prompt to a generative model (e.g., ChatGPT), inspects the returned artifact, optionally modifies the prompt several times, and then accepts one of the results or abandons this attempt~\cite{DBLP:conf/chi/GoloujehSM24}. 
This paradigm may work reasonably well when the user has a clear vision of what they want and can articulate highly specific prompts. 
For creative work, however, this paradigm introduces several challenges.

\looseness-1
First, users are presented with fully rendered artifacts right away. This can lead to premature convergence and design fixation, where users are anchored to initial outputs without exploring broad, divergent idea generation~\cite{DBLP:conf/chi/Wadinambiarachchi24,DBLP:conf/chi/SuhCMLX24}. 
In Finke et al.'s terms, this workflow eliminates the \emph{generative} process in creative work.

\looseness-1Second, once users decide to refine an initial output, they frequently struggle to communicate nuanced changes to generative models and often become fixated on a single or a few refined outputs~\cite{DBLP:conf/chi/Zamfirescu-Pereira23,DBLP:conf/chi/SubramonyamPPAS24,DBLP:conf/nips/YuZHF024,DBLP:conf/chi/WangHS0024}. There is a gap between the mental vision users have and the prompts they are able to write (i.e., the ``gulf of envisioning''~\cite{DBLP:conf/chi/Zamfirescu-Pereira23,DBLP:conf/chi/SubramonyamPPAS24}). Users typically only recognize what feels wrong or what they wish to adjust after seeing the output, yet they cannot easily translate these intentions into concrete textual instructions that the model can interpret reliably. For instance, a user may want the model to ``make the image feel more lively'' for a street scene. The model may execute this by adding bright saturated colors, and the user only realizes this is not what they intended after seeing the result, pushing them into a trial-and-error prompting loop. Moreover, users can also fixate on the model's initial refinement results. For example, the model may execute the prompt ``make the image feel more lively'' by changing the time of day, and the user feels this is roughly correct and accepts it. However, other possibilities, such as adding small groups of people, may remain unexplored. As a result, users may end up refining through local edits, reinforcing design fixation and leading to suboptimal convergence.

Third, creative work is often non-linear and involves iteratively moving between generating possibilities and refining, revising, or recombining them as their goals evolve. However, current chatbot interfaces often adopt a linear conversation structure, offering limited support for such non-linear creative processes. 

Based on this analysis, we propose three design goals (DGs):

\begin{itemize}[leftmargin=*]
    \item \textbf{DG1: Scaffold divergent generation of high-level ideas to reduce design fixation.}
    The system should support users in generating diverse high-level ideas before committing to an actual artifact. By doing so, the system aims to reduce premature convergence and design fixation.

    \item \textbf{\looseness-1DG2: Support convergent refinement to bridge the ``gulf of envisioning'' and avoid fixation.}
    The system should help users translate their intentions into concrete, actionable refinements and encourage further exploration of alternative refinements rather than fixation on a single ``good enough'' output.

    \item \looseness-1\textbf{DG3: Enable non-linear, iterative workflows that preserve context.}
    The system should support fluid movement between divergent and convergent modes, including branching and revisiting earlier ideas. Ideas and artifacts should be organized to allow users to revisit and build on them without losing history. 
\end{itemize}



\begin{figure}[t!]
    \centering
    \includegraphics[width=0.8\linewidth]{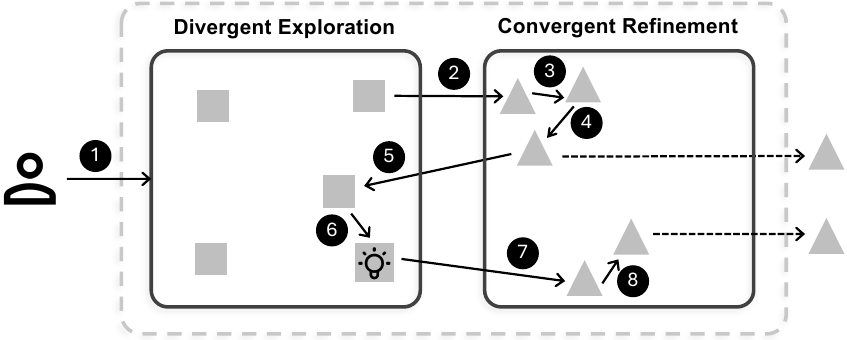}
    \caption{\looseness-1A generalization of Fig.~\ref{fig.teaser}, exemplifying a user trajectory when using our two-mode approach to solve a task. \graysquare{} denote an idea, and \graytriangle{} denote an artifact.}
    \Description{A generalized workflow figure showing users alternating between divergent ideation and convergent image refinement while preserving history across both modes.}
    \label{fig.system.overview}
\end{figure}

\section{System Design} \label{sec:method}

\looseness-1This section presents our system design addressing the above problems. Section~\ref{sec:method:overview} discusses the overall approach, and how we instantiate it in our system, \ours{}. Sections~\ref{sec:method:brainstorm} and \ref{sec:method:refine} describe two modes of \ours{}. Section~\ref{sec:method:validation} provides the technical validation of our system . We provide further implementation details in Appendix~\ref{appendix:implementation}.

\subsection{Approach Overview} \label{sec:method:overview}

\looseness-1We examine an approach grounded in the Geneplore model of creativity~\cite{finke1996creative} (see Fig.~\ref{fig.system.overview}).
Finke et al.'s Geneplore model of creativity describes creative cognition through two processes: \emph{generative} and \emph{exploratory}, which are related to the cognitive modes of divergent and convergent thinking~\cite{guilford1956structure,guilford1967nature}.
In the \emph{generative} process, users produce diverse high-level ``preinventive'' structures without strong goal control (i.e., divergent thinking). In the \emph{exploratory} process, these structures are interpreted and developed with respect to the specific task (i.e., convergent thinking). Based on this model, we examine a system design that comprises two corresponding modes: \firstmode{} and \secondmode{}. The \firstmode{} mode scaffolds users in the divergent generation of high-level ideas (DG1), while the \secondmode{} mode supports users in the convergent refinement of those ideas to match their objectives for the task (DG2). By allowing users to switch back and forth between the two modes without losing interaction history, the approach enables a non-linear, iterative workflow for creative tasks (DG3). We operationalize this approach through \ours{}, our web-based co-creation system, which serves as an instantiation of the approach for the domain of creative image generation. Next, we describe how the \firstmode{} and \secondmode{} modes are operationalized within \ours{}.

\subsection{\firstmode{} Mode} \label{sec:method:brainstorm}


\begin{figure*}[htbp!]
    \centering
    \renewcommand{\thesubfigure}{\roman{subfigure}}
    \includegraphics[width=0.99\linewidth]{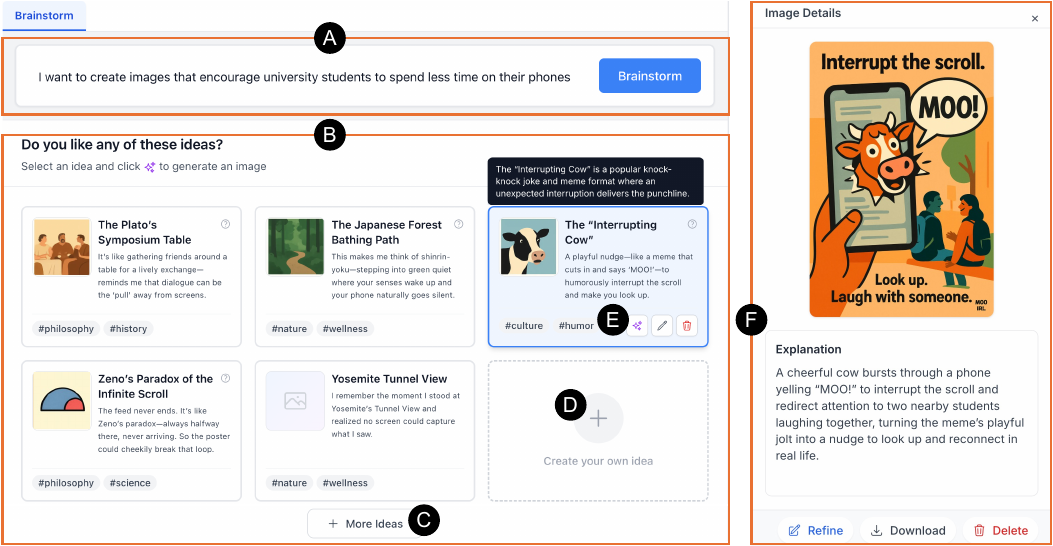}\\[-8pt]

    \begin{subfigure}[t]{0.70\linewidth}
        \caption{The interface of the \firstmode{} mode.}
        \label{fig.system.interface.brainstorm}
    \end{subfigure}%
    \hfill
    \begin{subfigure}[t]{0.28\linewidth}
        \caption{The interface of the image viewer.}
        \label{fig.system.interface.generation}
    \end{subfigure}

    \vspace{1em}
    \includegraphics[width=0.99\linewidth]{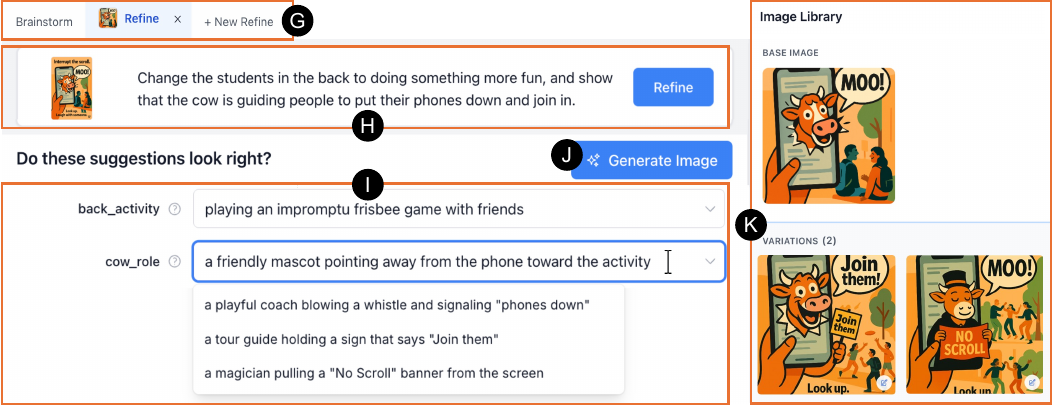}\\[-8pt]

    \begin{subfigure}[t]{0.95\linewidth}
        \caption{The interface of the \secondmode{} mode.}
        \label{fig.system.interface.refine}
    \end{subfigure}


    \caption{
    \looseness-1 \ours{}'s interfaces. 
    In the \firstmode{} mode, the user starts by \blackcircled{A} entering a prompt, after which \ours{} \blackcircled{B} generates a set of \emph{Idea Cards} with a title, thumbnail, description, background (shown on hover of the question mark), and category tags. The user can continue broadening the idea space by \blackcircled{C} clicking ``More Ideas'' or \blackcircled{D} manually creating their own idea card, for example, by creating ``Yosemite Tunnel View'' as a new idea inspired by ``The Japanese Forest Bathing Path.'' Upon selecting an idea, the user can \blackcircled{E} edit it with the \emph{pencil icon} or \blackcircled{F} generate an image from that idea with the \emph{spark icon}. 
    To refine an image, clicking the ``Refine'' button (Fig.~\ref{fig.system.interface}F) \blackcircled{G} opens a new refinement tab; multiple tabs can be opened in parallel. 
    In the \secondmode{} mode, the user can \blackcircled{H} submit a refinement prompt, after which the system \blackcircled{I} generates \emph{parameters} (whose meanings are shown on hover of the question mark) and \emph{options} that the user can select (options are also editable); the user can then \blackcircled{J} generate a new image variation. The \blackcircled{K} \emph{Image Library} shows the initial image alongside its refined variations, and any image can be further refined via the icon in its bottom-right corner. The trajectory in Fig.~\ref{fig.teaser} maps to these UI operations as follows: \blackcircled{1} $\rightarrow$ \blackcircled{A}\blackcircled{B}, \blackcircled{2} $\rightarrow$ \blackcircled{E}\blackcircled{F}, \blackcircled{3}\blackcircled{4} $\rightarrow$ \blackcircled{H}\blackcircled{I}\blackcircled{J}\blackcircled{K}, \blackcircled{5} $\rightarrow$ \blackcircled{G}, \blackcircled{6} $\rightarrow$ \blackcircled{D}, \blackcircled{7} $\rightarrow$ \blackcircled{E}\blackcircled{F}, and \blackcircled{8} $\rightarrow$ \blackcircled{H}\blackcircled{I}\blackcircled{J}\blackcircled{K}.
    }
    \label{fig.system.interface}
\end{figure*}

\looseness-1Within \ours{}, the \firstmode{} mode attempts to reduce idea fixation by making conceptual exploration the first action before any actual image artifact is generated. The interface of the \firstmode{} mode is shown in Fig.~\ref{fig.system.interface}.
For instance, suppose a user wants to create an image that encourages students to spend less time on their phones.
The user enters a task description (Fig.~\ref{fig.system.interface}A), for example, {``I want to create an image to encourage students to spend less time on their phones.''} \ours{} then populates an \emph{Idea Grid} with structured \emph{Idea Cards} (Fig.~\ref{fig.system.interface}B). Each idea card represents a distinct high-level conceptual idea, such as a metaphor, philosophical concept, or a cultural reference, rather than surface-level dimensional attributes such as color and style (DG1). Each idea card contains a title, an idea thumbnail, a brief description, category tags, and contextual background revealed on hover. This multi-component format lets users scan and compare conceptual directions at a glance, filtering by relevance without wading through unstructured text.

\looseness-1
Importantly, users are not passive recipients of these ideas. They can expand the idea grid with a steering prompt to generate additional ideas in a chosen direction (Fig.~\ref{fig.system.interface}C), create their own cards by entering a title and brief description (Fig.~\ref{fig.system.interface}D), or edit existing cards to shape the space toward their intent. They can click the \emph{spark icon} (Fig.~\ref{fig.system.interface}E) to generate an image from a selected idea; the resulting image viewer displays the image alongside an \emph{explanation} of how it fits the original task (Fig.~\ref{fig.system.interface}F).

\looseness-1
Generating ideas that are genuinely creative and diverse, rather than surface variations on the user's prompt, requires more than asking a generative model to ``be creative and diverse.'' Drawing on prior research showing that \emph{associative thinking}, the capacity to connect remote and seemingly unrelated concepts, reliably boosts creative output~\cite{beaty2023associative,mednick1962associative}, we explicitly prompt the model to adopt an associative-thinking strategy that draws connections from distant domains, such as artworks, mythology, historical events, and internet culture, among others~\cite{DBLP:journals/corr/abs-2405-06715}. This strategy produced the ``Interrupting Cow'' idea in Fig.~\ref{fig.system.interface}: the model connected ``disrupting a phone habit'' to the disruptive logic of the meme, yielding a concept the user may not be able to think of independently. 
An ablation study reported in Section~\ref{sec:results:rq2} confirms that associative-thinking prompting produces significantly higher idea diversity than non-associative-thinking prompting, validating this design decision.

\subsection{\secondmode{} Mode} \label{sec:method:refine}

\looseness-1
By generating an image for an idea they find promising, the user enters the \secondmode{} mode by selecting the image and clicking ``Refine'' (Fig.~\ref{fig.system.interface}F), which opens a dedicated \emph{Refine Tab} alongside the Brainstorm Tab (Fig.~\ref{fig.system.interface}G). The challenge at this point is that users may recognize that something in the image needs to change, but cannot easily translate that intuition into a precise prompt~\cite{DBLP:conf/chi/SubramonyamPPAS24,DBLP:conf/chi/Zamfirescu-Pereira23}. The model may interpret the user's underspecified instructions arbitrarily, leading to trial-and-error and dissatisfaction.
\looseness-1 \ours{} addresses this by acting as a \emph{translator} (DG2): the user states a refinement intent in their prompt (Fig.~\ref{fig.system.interface}H, e.g., {``Change the students in the back to doing something more fun.''}), and the system synthesizes a \emph{Sketch}, a parametric function that decomposes that intent into named semantic parameters and pre-populates each parameter with concrete, context-aware dropdown options (Fig.~\ref{fig.system.interface}I). For instance, the system infers \texttt{cow\_role} as a relevant parameter for the user's refinement intent and offers options such as ``a friendly mascot,'' ``a playful coach,'' and ``a tour guide,'' each representing a distinct interpretation of the user's intent. These options can also be customized and edited by the user. 
For each parameter, the pre-filled default option is intentionally designed to best align with the user's refinement intent.

\ours{} also supports iterative refinement. 
After exploring a set of image variations (Fig.~\ref{fig.system.interface}K), users can select any individual variation and initiate a further round of refinement, building successive chains of targeted adjustments. 
Users can also modify the refinement prompt to generate a new set of parameters and options, open multiple parallel Refine Tabs to develop different image directions.
\ours{} also enables non-linear workflows: users can return to the Brainstorm Tab (i.e., \firstmode{} mode) to explore new conceptual ideas, all without losing the context of prior attempts (Fig.~\ref{fig.system.interface}G; DG3).

\subsection{Technical Validation} \label{sec:method:validation}

We validated the technical implementation of \ours{}.
To validate the \firstmode{} mode, we randomly sampled 100 brainstorming and ideation prompts from the Infinity-Chat dataset~\cite{DBLP:journals/corr/abs-2510-22954}. The results show that the system generated the idea grid in $14.0$s on average, completed the full generation of the idea grid and thumbnails in $33.5$s on average, and successfully generated all thumbnails for $97\%$ of prompts.
To validate the \secondmode{} mode, we randomly sampled $100$ pairs of (image, refinement prompt) from the MagicBrush dataset~\cite{DBLP:conf/nips/ZhangMCSS23}. The results show that sketch generation took $9.01$s on average; the generated sketches were $100\%$ syntactically valid. Full setup and more results are reported in Appendix~\ref{appendix:technical_validation}.


\section{User Study and Analysis Setup} \label{sec:user_study}

We evaluate \ours{} with the following research questions (RQs):

\begin{itemize}[leftmargin=*]
    \item \textbf{RQ1 (Experience and Creative Outputs):} How does \ours{}'s two-mode workflow affect users' co-creation experience and the quality of their creative outputs?

    \item \textbf{RQ2 (Brainstorming and Exploration):} How does scaffolding divergent ideation through the \firstmode{} mode affect early-stage exploration?

    \item \textbf{RQ3 (Refinement and Control):} How does scaffolding convergent refinement through the \secondmode{} mode affect users' refinement process?

    \item \textbf{RQ4 (Learning):} Does using \ours{} lead to self-reported learning, and if so, what knowledge or strategies do users learn?
\end{itemize}

\looseness-1
To address these RQs, we conducted a within-subjects study ($N=24$) comparing \ours{} against ChatGPT, one of the most widely used text-to-image interfaces for general users. 
Because the two systems differ along multiple interface and workflow dimensions, this comparison evaluates the overall approach rather than isolating individual components. 
Both systems used GPT-5~\cite{gpt5} for text generation and OpenAI's image generation models throughout.\footnote{Both systems initially used \texttt{gpt-image-1}~\cite{openai2024gptimage1}; mid-study, ChatGPT upgraded to \texttt{gpt-image-1.5}~\cite{openai2024gptimage1.5}, and we synchronized \ours{} accordingly. Each model version was used by 12 participants, equally distributed across conditions. We did not observe substantial differences in results between the two model-version groups.} To ensure a fair comparison, generating multiple images simultaneously was not allowed in either system.
Participants were asked to create poster images promoting a specific behavior. We chose the task of creating poster images because it is a widely accessible and non-professional activity requiring both conceptual thinking and creativity. 
To reduce practice effects, we designed three slightly different tasks of similar difficulty: (A) ``Spend less time on phones,'' (B) ``Spend more time outdoors,'' and (C) ``Take care of your mind.'' The complete task instructions are provided in Appendix~\ref{appendix:userstudy_task_instruction} (Fig.~\ref{fig.experiments.tasks}). 
Each participant completed two image co-creation sessions, one with \ours{} and one with ChatGPT, with a different task for each session. 
This task assignment followed a Balanced Incomplete Block Design~\cite{lazar2017research} to ensure equal coverage. This design resulted in 12 counterbalancing sequences (3 task pairs $\times$ 2 task orders $\times$ 2 system orders). 
Accordingly, we recruited 24 participants, with two participants randomly assigned to each sequence (see Appendix~\ref{appendix:userstudy}, Table~\ref{tab:userstudy_bibd}).

\subsection{Participants} \label{sec:userstudy:participants}

\looseness-1
We recruited 24 participants (10 female, 13 male, 1 non-binary) from the local university via mailing lists and word of mouth (aged 24--34 years, $M=28.2$, $SD=2.8$). Since the university is technology-focused, the participants were primarily from the computer science/information technology domain (21/24), with two from other engineering domains and one from the educational domain. None of them had formal training in art or design, although 12/24 reported self-taught or informal experience (7 for less than 1 year, 5 for 1--3 years). All had heard of AI-based image tools, and 22/24 had prior hands-on experience (e.g., ChatGPT: 22, Canva: 12, DALL-E: 10, Gemini: 7, Midjourney: 7, Stable Diffusion: 4). Creative activity frequency varied (daily: 1; weekly: 5; monthly: 6; a few times per year: 10; never: 2). Detailed participant demographics are provided in Appendix~\ref{appendix:userstudy} (Table~\ref{tab:userstudy_demographics}).
Each participant received a \texteuro{20} voucher and a reusable bag. The study was approved by the local Institutional Review Board.


\begin{figure*}[t!]
    \centering
    \includegraphics[width=\textwidth]{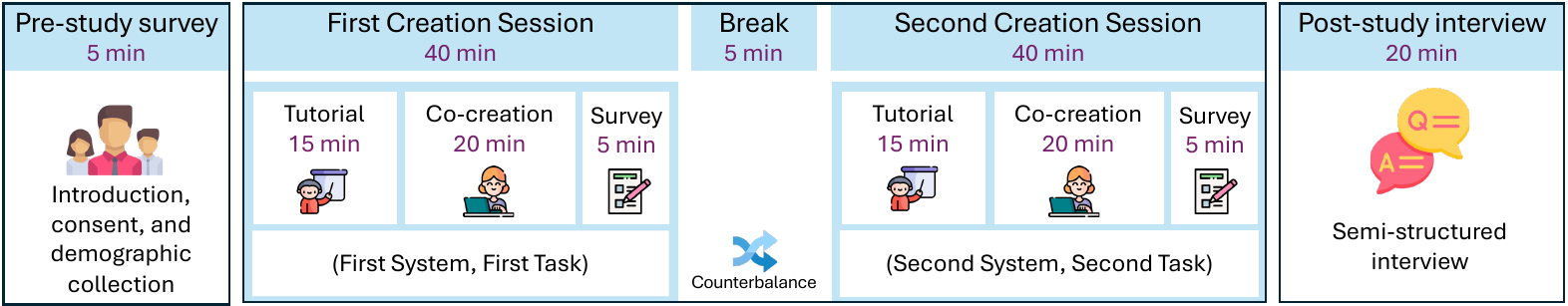}
    \caption{\textbf{User Study Procedure} (figure adapted from~\cite{10.1145/3706598.3713375}).}
    \label{fig.experiments.procedure}
\end{figure*}

\subsection{Study Procedure} \label{sec:userstudy:procedure}

\looseness-1
The study procedure is illustrated in Fig.~\ref{fig.experiments.procedure}.
The study lasted about 1 hour and 50 minutes.
Participants first signed a consent form and completed a questionnaire on demographics, prior AI experience, and creative/design experience. 
The study comprised two image co-creation sessions. Each session began with onboarding comprising a standardized system tutorial (participants could end early if they felt ready) and free exploration (up to 15 minutes total).
Participants then worked on their assigned task with screen interactions recorded. At the end of each session, participants downloaded images they considered ready to print as a poster and completed a post-task survey. After a break, participants completed the second session with a different task and the other system, following the same procedure.
Finally, participants completed a semi-structured interview covering each system's workflow, creative exploration, perceived learning, and intentions for future use (see Appendix~\ref{appendix:userstudy}). Interviews were audio-recorded and transcribed for thematic analysis.

\subsection{Data Collection and Analysis} \label{sec:userstudy:data_analysis}

We collected quantitative data from surveys, screen recordings, and downloaded images, and qualitative data from semi-structured interviews. For all statistical comparisons, we used the Wilcoxon signed-rank test with $\alpha = 0.05$~\cite{wilcoxon1992individual}.

\looseness-1\emph{Post-task Surveys.} 
Post-task surveys included the unweighted Creativity Support Index (CSI)~\cite{DBLP:journals/tochi/CherryL14} (items rated 0--10 and summed per dimension to a 0--20 range; for RQ1) and UMUX-Lite~\cite{DBLP:conf/chi/LewisUM13} for system usability (items rated 1--7, with the overall score mapped to 0--100; for RQ1). Additionally, we included a self-reported learning item ``While using this system, I learned something new that was previously unknown to me'' (item rated 1--7 indicating strongly disagree to strongly agree; for RQ4). This item is followed by an optional open-ended question ``If you agree, please write a few keywords or short phrases about what you learned'' (for RQ4). We used the unweighted CSI without paired comparisons to reduce participant burden~\cite{10.1145/3698061.3726933} and omitted the Collaboration dimension because the study was non-collaborative, consistent with prior work~\cite{DBLP:conf/chi/SuhCMLX24,DBLP:conf/uist/SuhZL22}. Open-ended responses about learning were independently labeled by two researchers and discussed to reach consensus (see Appendix~\ref{appendix:learning_labels} for more details about the coding scheme).

\looseness-1\emph{Behavioral Metrics from Screen Recordings.}
We derived four behavioral metrics: (i) \textit{Image Clusters} (both systems; for RQ2) --- the number of distinct conceptual ideas explored, where each cluster comprises an initial image and its refinement variations; (ii) \textit{Refinement Prompts per Cluster} (both systems; for RQ3) --- the average number of refinement prompts per image cluster; (iii) \textit{Non-Default Option Adoption} (\ours{} only; for RQ3) --- the proportion of refinements where participants changed at least one default option; and (iv) \textit{Brainstorm-First Adoption} (ChatGPT only; for RQ4) --- whether a participant's first ChatGPT prompt requested brainstorming ideas rather than a direct image, compared across system-order groups to assess whether prior \ours{} experience transferred to ChatGPT.

\looseness-1\emph{Downloaded Images.}
We evaluated images downloaded by participants that they considered ready to print as posters. We assessed them on \emph{Fluency}, \emph{Diversity}, \emph{Novelty}, and \emph{Usefulness}~\cite{runco2012standard,harvey2023toward,doi:10.1126/sciadv.adn5290} (for RQ1). 
We operationalized \emph{Fluency} as the number of images downloaded per participant~\cite{runco2012standard}. \emph{Diversity} was measured as average pairwise cosine distance between image embeddings calculated using CLIP ViT-bigG-14~\cite{DBLP:conf/nips/SchuhmannBVGWCC22,DBLP:conf/chi/KumarVJA25,doi:10.1126/sciadv.adn5290}; three participants who downloaded only one image for both systems were excluded, yielding $N=21$. 
We operationalized \emph{Novelty} as the perceived originality of a poster's idea and \emph{Usefulness} as the effectiveness of a poster in promoting the intended behavior.
\textit{Novelty} and \textit{Usefulness} were rated by five independent raters (not authors) on a 5-point Likert scale~\cite{amabile1982social,doi:10.1126/sciadv.adn5290}.  
Four of the five raters had also participated in the user study, but ratings were blind, and each participant's own images constituted only a small fraction (less than 5\%) of the image pool.
Per-image scores were averaged first across raters, then across images per (participant, system) pair, and finally across participants to yield per-system scores. 
We aggregated ratings across five raters, yielding moderate inter-rater reliability (ICC(2,5): Novelty = 0.61, Usefulness = 0.60)~\cite{koo2016guideline}.
More details are provided in Appendix~\ref{appendix:image_annotation}.

\looseness-1
\emph{Ablation: Associative-Thinking Prompting on Idea Diversity.}
For RQ2, we examine associative-thinking prompting in \firstmode{} mode by comparing two conditions: (i) an \emph{associative-thinking} prompt that specifically instructs the model to generate ideas by drawing on remote connections across domains, and (ii) a \emph{non associative-thinking} prompt that simply instructs the model to generate creative and diverse ideas (see Appendix~\ref{appendix:prompts}). After the study, using the 24 initial prompts participants entered into \ours{}, we queried GPT-5 under both conditions three times each (9 ideas per run, matching \ours{}'s configuration), computed embedding-based diversity as the average pairwise cosine distance between CLIP ViT-bigG-14 encodings per run, and averaged the scores across runs. 

\looseness-1
\emph{Thematic Analysis.}
We analyzed interview transcripts using thematic analysis~\cite{clarke2017thematic,Braun01012006} (for RQ1--4). Two researchers independently coded transcripts from two randomly selected participants, met to discuss and refine codes, then each coded half of the remaining transcripts before meeting again to consolidate codes into themes.


\section{Results} \label{sec:results}

This section presents our findings by RQs. 
In all figures, error bars represent bootstrapped 95\% confidence intervals, and $^{*}p < 0.05$, $^{**}p < 0.01$, and $^{***}p < 0.001$.


\begin{figure*}[t!]
    \centering
    \includegraphics[width=\linewidth]{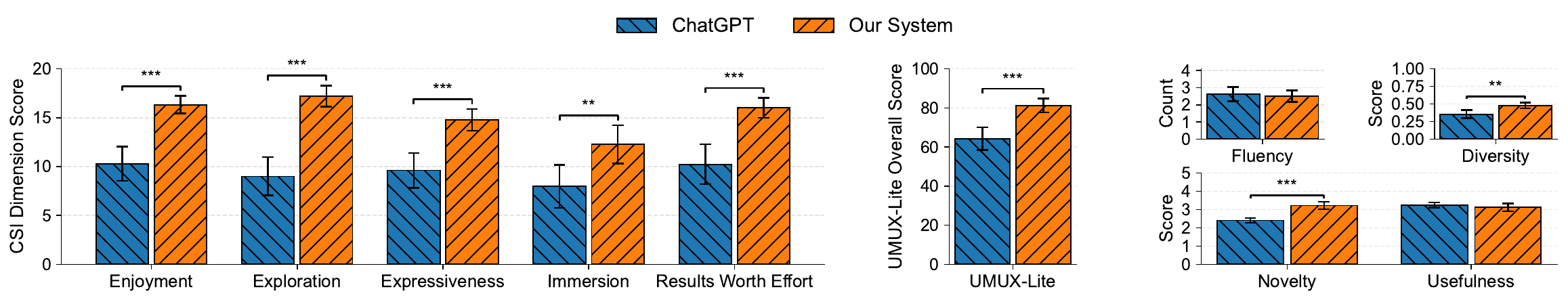}
    \\[-12pt]

    \begin{subfigure}[t]{0.52\linewidth}
        \vspace{-4pt}
        \caption{Creativity Support Index (CSI)}
        \label{fig:rq1_csi}
    \end{subfigure}%
    \hfill
    \begin{subfigure}[t]{0.12\linewidth}
        \vspace{-4pt}
        \caption{UMUX-Lite}
        \label{fig:rq1_system_usability}
    \end{subfigure}%
    \hfill
    \begin{subfigure}[t]{0.25\linewidth}
        \vspace{-4pt}
        \caption{Final Images}
        \label{fig:rq1_final_images}
    \end{subfigure}
    \\[-8pt]
    \caption{Results for RQ1. (a) Creativity Support Index (CSI) scores across five dimensions: \ours{} scored significantly higher than ChatGPT on all dimensions (all $W < 30.0$, all $p < 0.002$). (b) System usability (UMUX-Lite): \ours{} scored significantly higher than ChatGPT overall ($M = 81.25$ vs.\ $64.24$; $W = 17.0$, $p < 0.001$). (c) Final downloaded image quality across four dimensions: \ours{} produced significantly more novel images ($M = 3.22$ vs.\ $2.41$; $W = 0.0$, $p < 0.001$) and more diverse image sets ($M = 0.48$ vs.\ $0.36$; $W = 26.0$, $p = 0.001$), while fluency and usefulness were not significantly different across systems.}
    \label{fig:rq1}
\end{figure*}


\begin{figure*}[t!]
    \centering
    \begin{subfigure}[t]{0.246\linewidth}
        \centering
        \includegraphics[width=\linewidth]{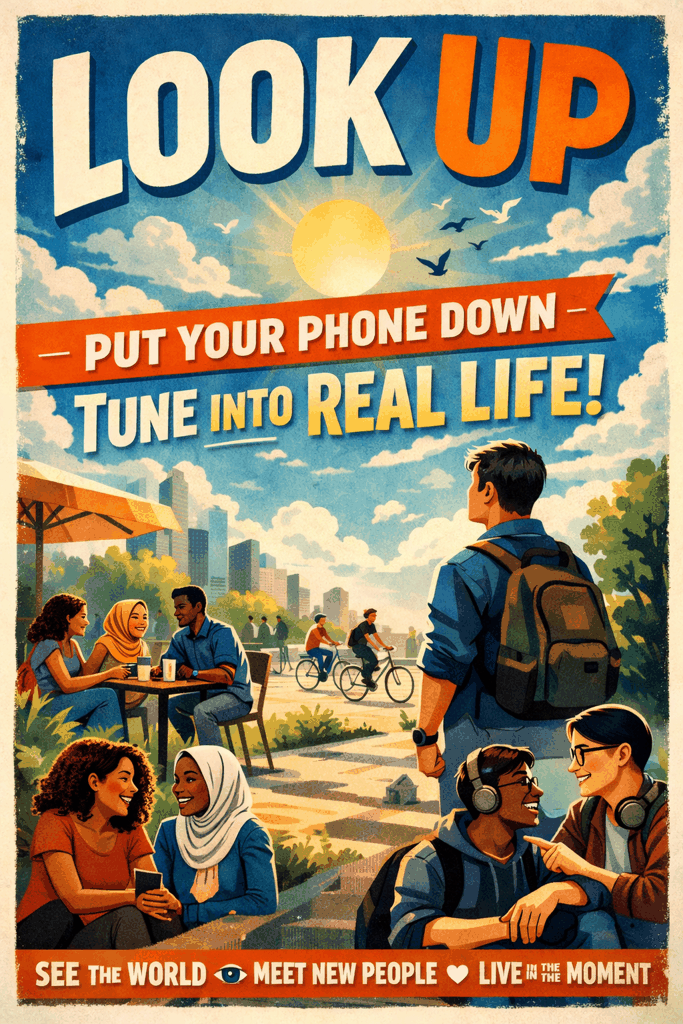}
        \caption{ChatGPT (best)}
    \end{subfigure}
    \hfill
    \begin{subfigure}[t]{0.246\linewidth}
        \includegraphics[width=\linewidth]{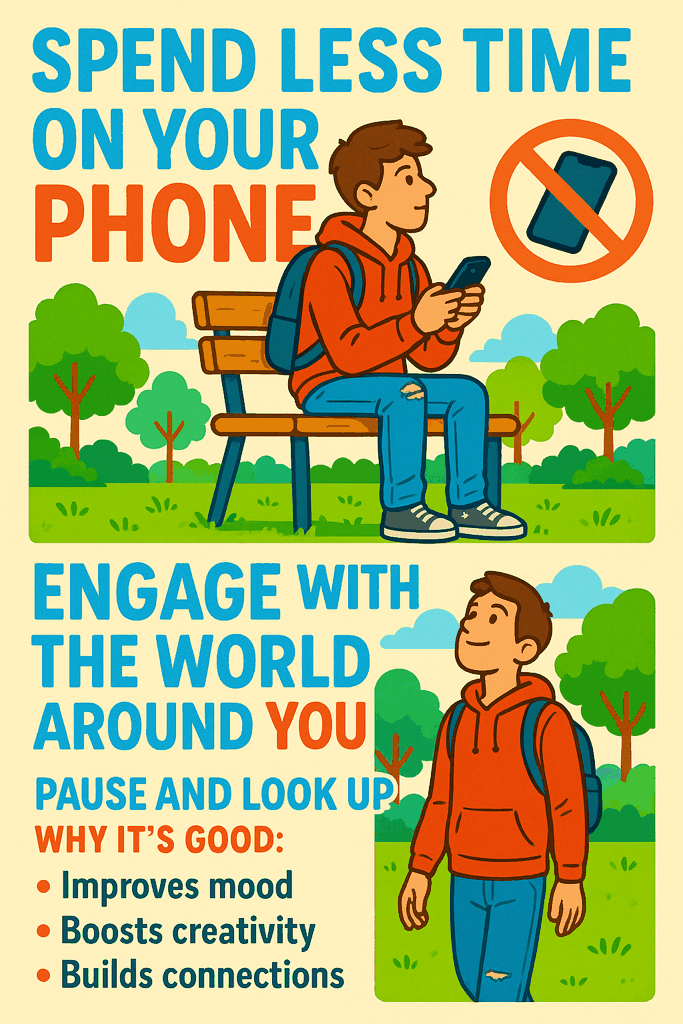}
        \caption{ChatGPT (median)}
    \end{subfigure}
    \hfill
    \begin{subfigure}[t]{0.246\linewidth}
        \centering
        \includegraphics[width=\linewidth]{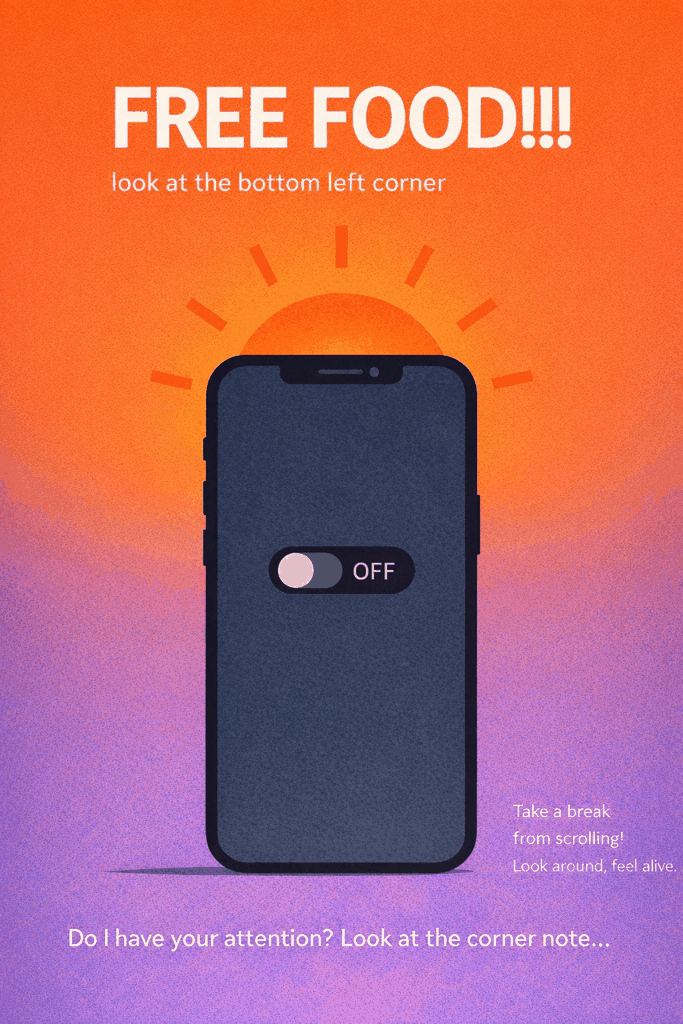}
        \caption{\ours{} (best)}
    \end{subfigure}
    \hfill
    \begin{subfigure}[t]{0.246\linewidth}
        \includegraphics[width=\linewidth]{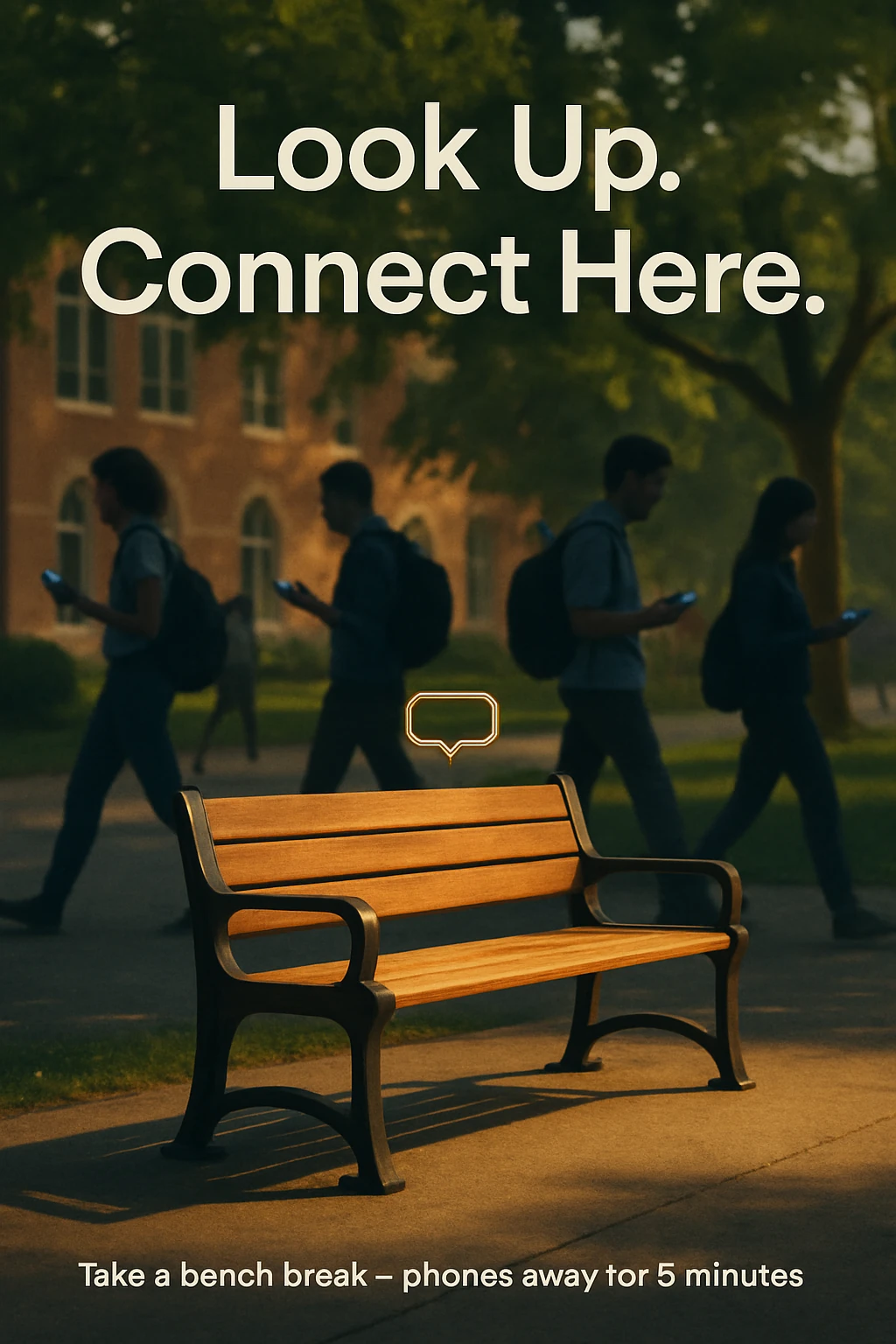}
        \caption{\ours{} (median)}
    \end{subfigure}
    
    \caption{Example final posters created during the study for the task ``Spend Less Time on Phones.'' Posters (a) and (b) were created with ChatGPT, and (c) and (d) were created with \ours{}. For each system, we show the highest overall-scoring poster and a median overall-scoring poster, where the overall score is the sum of the novelty and usefulness scores. The (novelty, usefulness) scores for posters (a)-(d) are $(2.4, 3.8)$, $(2.0, 3.8)$, $(4.8, 3.8)$, $(3.2, 3.4)$, respectively. Poster (c) achieved a high novelty score (4.8/5) by first capturing attention with ``FREE FOOD!!!,'' then guiding viewers across the layout from the left corner to the right corner, where the key message appears: ``Take a break from scrolling! Look around, feel alive.'' As viewers follow these cues, their attention shifts away from the central phone, subtly mirroring the act of spending less time on it. This staged redirection makes the message both novel and memorable.}
    \label{fig:rq1_examples}
\end{figure*}

\subsection{RQ1: Experience and Creative Outputs} \label{sec:results:rq1}

\looseness-1This section presents results on RQ1: How does \ours{}'s two-mode workflow affect users' co-creation experience and the quality of their creative outputs? (see Fig.~\ref{fig:rq1})\footnote{\looseness-1We did not observe significant differences in quantitative results across participants' backgrounds (e.g., design experience) within each system in this and subsequent RQs.}

\looseness-1
\ours{} scored significantly higher than ChatGPT for creativity support (Fig.~\ref{fig:rq1_csi}) and system usability (Fig.~\ref{fig:rq1_system_usability}), and participants produced images with significantly higher \textit{Novelty} and \textit{Diversity} scores with \ours{} (Fig.~\ref{fig:rq1_final_images}).
Fig.~\ref{fig:rq1_examples} shows example final posters created with the two systems.
In interviews, most participants explicitly expressed a preference for \ours{} over ChatGPT for creative image generation. Five participants noted that ChatGPT remains useful when they already have a clear vision of what they want: ``\emph{I would actually prefer \ours{}, unless I really have a very specific idea in mind; in that case, I could just use ChatGPT. But I feel like these cases are actually very rare.}'' (P10).

\looseness-1
Nine participants noted that ChatGPT's single-threaded history made it difficult to track or revisit earlier ideas, while \ours{}'s tab-based structure enabled them to explore multiple directions in parallel. 
P2 said, ``\emph{I can go back and forth between ideas, keep track of them, and explore each of them further as I want.}''
The tab structure also repurposed an unavoidable bottleneck: most participants used the waiting time during image generation to explore other ideas in parallel. As P1 noted, ``\emph{I could brainstorm parallelly multiple different ideas while waiting for some ideas to be implemented}''.

\looseness-1
Perceived agency and ownership varied across participants. P1 and P8 reported reduced agency with \ours{}, describing their role as guiding and curating system-generated ideas rather than authoring the core concept themselves. In contrast, P16 and P18 felt \ours{} gave them \textit{more} control than ChatGPT, citing its structured options as enabling finer-grained authorial direction. P10 and P21 reported equally low ownership across both systems, viewing AI-generated images as inherently not their own regardless of interface. Several other participants (P4, P12, P19) framed the experience as co-creation; they acknowledged that the system contributed ideas but did not frame this as a loss of agency.



\begin{figure}[t!]
\centering
\includegraphics[width=0.8\columnwidth]{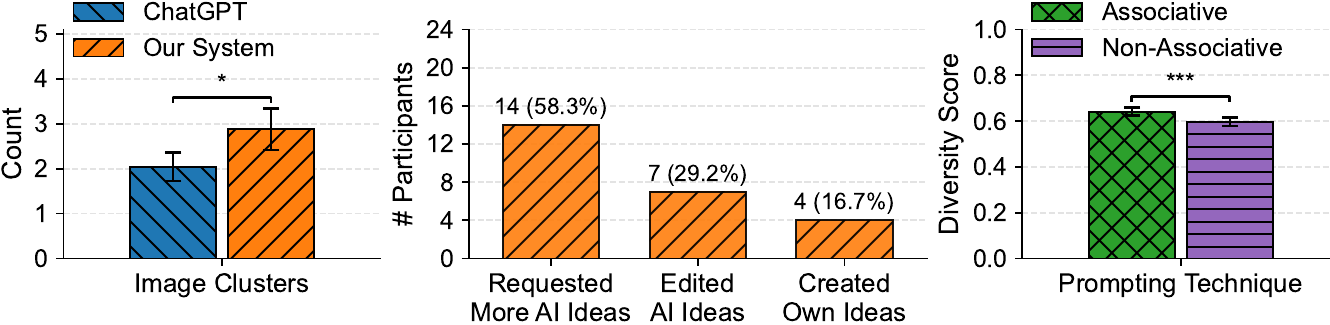}
\caption{Results for RQ2. Left: participants had significantly more image clusters with \ours{} than with ChatGPT ($M = 2.88$ vs.\ $2.04$; $W = 33.0$, $p = 0.021$). Middle: participant engagement with the \firstmode{} mode in \ours{}. Right: associative-thinking prompting used in \firstmode{} mode produced significantly higher idea diversity than non associative-thinking prompting ($W = 24.0$, $p < 0.001$).}
\label{fig:rq2}
\end{figure}

\subsection{RQ2: Brainstorming and Exploration} \label{sec:results:rq2}

This section presents results on RQ2: How does scaffolding divergent ideation through the \firstmode{} mode affect early-stage exploration? (see Fig.~\ref{fig:rq2})

Participants examined significantly more distinct conceptual ideas with \ours{} than with ChatGPT (Fig.~\ref{fig:rq2}, left). They also actively engaged with the ideas in \ours{} beyond passively selecting them (Fig.~\ref{fig:rq2}, middle). 
In interviews, 13 participants described \ours{}'s \firstmode{} mode as helping them get started:
``\emph{The brainstorm part is crucial; when you want to create something, at the first stage, you 
may not know what you really want}'' (P14). 

Beyond helping participants get started, nine participants reported surprise at the surfaced diverse, cross-domain ideas. P1 recalled: ``\emph{I would never have thought of it in that direction. But just the fact that it showed me this did help me go down another rabbit hole.}'' P3 noted encountering the Pomodoro Technique reframed as a well-being tool: ``\emph{I knew the technique, but I wouldn't have thought of it while thinking about this.}'' This cross-domain reach was enabled in part by our associative-thinking prompting strategy, which produced significantly higher idea diversity than non associative-thinking prompting (Fig.~\ref{fig:rq2}, right). 

\looseness-1
Beyond idea diversity, nine participants described \ours{}'s structured idea cards as reducing evaluation effort by enabling rapid scanning and comparison. 
P5 noted that in ChatGPT ``\emph{the different ideas are kind of lost in the text, so it's hard to explore different ideas}'', whereas in \ours{} ``\emph{you see small visuals with a very short description, it's very easy to distinguish between ideas.}''

\subsection{RQ3: Refinement and Control} \label{sec:results:rq3}

\begin{figure}[t]
\centering
\includegraphics[width=0.5\columnwidth]{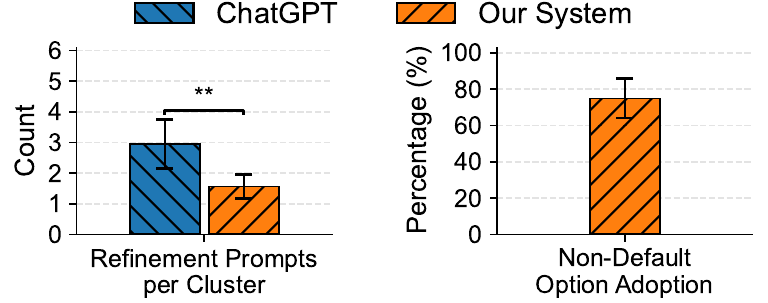}
\\[-4pt]
\caption{\looseness-1Results for RQ3. Left: participants made significantly fewer refinement prompts per image cluster with \ours{} than with ChatGPT ($M = 1.56$ vs.\ $2.94$; $W = 52.0$, $p = 0.004$). Right: participants used non-default options in $74.8\%$ of refinements on average ($SD=27.40\%$) when using \ours{}, indicating that alternative options were actively used.}
\label{fig:rq3}
\end{figure}

This section presents results on RQ3: How does scaffolding convergent refinement through the \secondmode{} mode affect users' refinement process? (see Fig.~\ref{fig:rq3})

Participants made significantly fewer refinement prompts per cluster with \ours{} than with ChatGPT (Fig.~\ref{fig:rq3}, left). 10 of 24 participants described ChatGPT's free-form refinement as unpredictable: it changed elements they intended to preserve (4/24), ignored targeted instructions (5/24), or required them to re-describe the entire image from scratch (4/24). 
P9 noted that ``\emph{[with ChatGPT] when I try to refine things it sometimes just randomly changes things},'' while P12 reported having to add explicit constraints like ``\emph{do not change the previous elements, only add}'' to prevent unintended overwrites. 
In contrast, 11 of 24 participants described \ours{}'s structured refinement options as making the refinement process easier and more controlled than free-text prompting.
P4 observed that the options served a dual communicative function: ``\emph{it doesn't just facilitate the system understanding what the user wants; it also helps the user understand what they want the system to do.}''

\looseness-1
The refinement options also surfaced new modification directions. As shown in Fig.~\ref{fig:rq3} (right), participants used non-default refinement options in 74.8\% of refinements rather than simply accepting the defaults. 7 of 24 participants reported that \ours{}'s refinement options introduced directions they would not have considered. 
P16 described being surprised by a suggestion they had not anticipated: ``\emph{I input some Mexican-style art, and it actually suggested an Aztec style. I didn't know that Mexican and Aztec things were kind of similar to each other. That was one idea I didn't know, and I felt like it was pretty good, so I went for it.}''

\looseness-1
However, for \ours{}, refinement fidelity and interaction flexibility were the primary concerns raised by participants. Four participants (P5, P8, P10, P22) reported that refined images did not always faithfully reflect their selected options; 
P8 noted the system ``\emph{sometimes fails to take these suggestions into consideration in the actual image.}'' 
Beyond fidelity, three participants (P1, P12, P20) desired direct manipulation of the image to complement the dropdown options, and two (P9, P12) expressed a desire for side-by-side comparison to more easily detect differences between generated images.


\subsection{RQ4: Learning} \label{sec:results:rq4}

\begin{figure*}[t!]
    \centering
    \includegraphics[width=\linewidth]{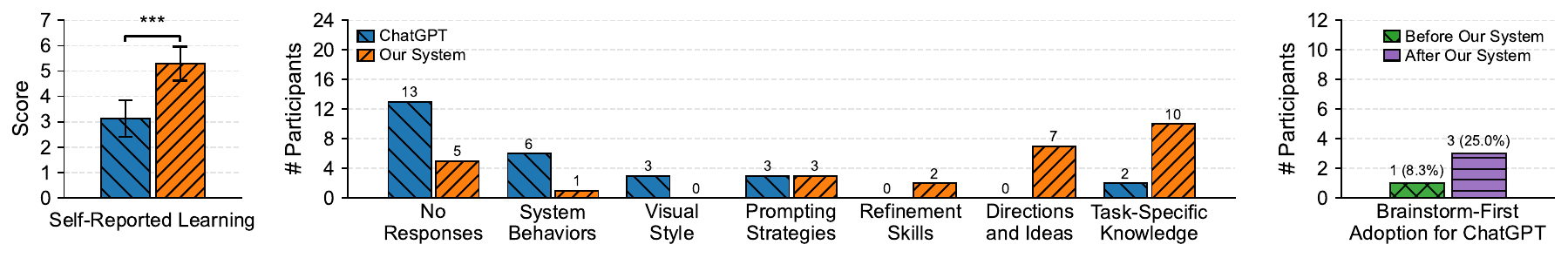}
    \caption{\looseness-1 Results for RQ4. Left: participants reported significantly higher self-reported learning with \ours{} than with ChatGPT ($M = 5.29$ vs.\ $3.12$; $W = 21.0$, $p < 0.001$). Middle: distribution of learning types from responses to the open-ended question about learning, showing that ChatGPT more often led to system learning (e.g., System Behaviors), whereas \ours{} more often led to task learning (e.g., Task-Specific Knowledge). Right: transfer of the brainstorm-first strategy to subsequent ChatGPT use, comparing participants who used ChatGPT before versus after \ours{}; after using \ours{}, more participants adopted a brainstorm-first workflow in ChatGPT, providing a preliminary signal that using \ours{} may help them internalize this workflow.}
    \label{fig:rq4}
\end{figure*}

This section presents results on RQ4: Does using \ours{} lead to self-reported learning, and if so, what knowledge or strategies do users learn? (see Fig.~\ref{fig:rq4})

\looseness-1
As shown in Fig.~\ref{fig:rq4} (left), participants reported significantly higher self-reported learning with \ours{} than with ChatGPT. 
Fig.~\ref{fig:rq4} (middle) shows the distribution of self-reported learning types. 13 of 24 ChatGPT participants did not provide any response, compared to 5/24 with \ours{}. Note that participants were instructed to provide responses if they \emph{agreed} that they had learned something. Among those who did report learning, ChatGPT participants more often reported learning about System Behaviors (6) and Prompting Strategies (3). For instance, P19 learned about a copyright restriction: ``\emph{I cannot use a superhero figure for image creation}.'' P14 learned that ``\emph{ChatGPT needs very clear and detailed instructions}.'' In contrast, \ours{} participants primarily reported learning of Task-Specific Knowledge (10) and New Directions and Ideas (7). See Appendix~\ref{appendix:learning_labels} for detailed label definitions and example responses.

\looseness-1
Beyond self-reported learning, we observed that a few participants, who used \ours{} in the first creation session, carried over the ``brainstorm-first'' workflow to ChatGPT in the second session (see Fig.~\ref{fig:rq4}, right).
During interviews, eight participants expressed an intention to apply this workflow in future tasks. P14 said, ``\emph{Maybe I will first ask: give me some options, rather than give them a prompt and directly an image}.'' P12 went further: ``\emph{Maybe I can create my own custom GPT that simulates this guidance. The prompts could be structured like this: first, when the user inputs tasks and requirements, it should verify them. Second, it should brainstorm and give potential ideas.}''


\section{Discussion} \label{sec:discussions}
This section discusses our findings and highlights the broader implications for future research on human-AI co-creation systems.

\subsection{Experience and Creative Outputs}

Quantitative results showed that \ours{} outperformed ChatGPT in perceived creativity support and system usability, and that with \ours{}, participants produced final images that were significantly more novel and diverse. Qualitative findings further showed a clear preference for \ours{}'s workflow, particularly its separation of divergent exploration and convergent refinement. At the same time, some participants noted that ChatGPT remains useful when they already have a clear vision of what they want. Perceived agency and ownership remained mixed across participants.

This preference for separating divergent exploration from convergent refinement directly relates to the challenge of design fixation in creative tasks~\cite{jansson1991design,DBLP:conf/chi/Wadinambiarachchi24}. In content generation, chatbot interfaces often produce a fully rendered artifact before any exploration has taken place, leading to premature convergence and design fixation~\cite{DBLP:conf/chi/Wadinambiarachchi24,jansson1991design}. \ours{} addresses this issue by structurally introducing scaffolded divergent exploration before any artifact is generated, thereby avoiding premature commitment to early outputs. In addition, \ours{} provides explicit process scaffolding for the two core cognitive processes (i.e., divergent and convergent thinking) that are central to creativity~\cite{wallas1926art,guilford1956structure,finke1996creative}.
Our results also align with recent work showing that structured exploration scaffolding helps users avoid design fixation~\cite{DBLP:conf/chi/SuhCMLX24} and that exploring multiple alternatives before committing leads to more divergent and higher-quality outcomes~\cite{DBLP:journals/tochi/DowGKSSK10}. 
However, the value of process scaffolding depends on the situation. P10 noted, ``\emph{I would actually prefer \ours{}, unless I really have a very specific idea in mind; in that case, I could just use ChatGPT.}'' This is broadly consistent with Horvitz's mixed-initiative framework~\cite{horvitz1999principles}, which emphasizes adapting system initiative based on factors such as user goal clarity, uncertainty, and the costs of interruption.

For system designers, these findings suggest that the ``chatbot'' paradigm may be ill-suited for creative tasks. Instead, generative tools should employ explicit process scaffolding to guide users through distinct stages of divergent and convergent thinking~\cite{guilford1967nature,runco1991divergent,cropley2006praise}. 
However, designers must navigate this shift with caution. 
First, users vary in how much scaffolding they need. Future systems should adapt scaffolding to users' current needs, supporting broad exploration when appropriate while allowing users to move quickly to execution when they are ready~\cite{DBLP:conf/chi/DhillonMLGZR24}. 
Second, introducing scaffolding can potentially reshape users' sense of agency and ownership~\cite{DBLP:journals/tochi/DraxlerWLHSBW24,salma2025designing}. Systems should therefore frame AI outputs as ``incomplete suggestions'' rather than ``complete solutions,'' ensuring that users retain responsibility for crucial decisions.

\subsection{Scaffolding Divergent Exploration}

Our results show that participants examined significantly more distinct conceptual ideas with \ours{} than with ChatGPT. Participants also expressed surprise at the diverse, cross-domain ideas surfaced through associative-thinking prompting. An ablation study demonstrates that this prompting strategy produced significantly more diverse ideas than simply asking the model to generate creative and diverse ideas.

These findings suggest that, in creative work, explicitly surfacing ideas is useful: although users can request ideas in a chatbot interface, they often do not explore broadly unless the system explicitly supports it. This is consistent with prior HCI work showing that dimensional reasoning about the design space does not arise naturally, motivating explicit scaffolding during early-stage divergent thinking~\cite{DBLP:conf/chi/SuhCMLX24}.
Our scaffolding for divergent exploration is implemented by surfacing diverse, cross-domain ideas through associative-thinking prompting. By doing this, our system shifts the user's role from independently generating ideas to evaluating, selecting, and building on them. When users encounter ideas they likely would not have produced on their own, they often experience surprise. 

Together, these findings suggest that simply asking a generative model to be creative or diverse is not enough. Instead, designers should employ explicit mechanisms for idea generation. 
Associative-thinking prompting is one strategy, but other prompting strategies may also help, such as compositional combination (blending familiar elements into novel hybrids), analogy formation (mapping structures across domains), and domain-specific abstraction (extracting reusable high-level patterns).
Additionally, recent work has raised concerns about homogeneity in open-ended language-model generation, showing both intra-model repetition and inter-model homogeneity~\cite{DBLP:journals/corr/abs-2510-22954}. Related work has further shown that LLM-based creativity support can homogenize ideas across users at the population level~\cite{DBLP:conf/candc/AndersonSK24}. Therefore, idea-generation mechanisms should be designed not only to broaden exploration for a single user, but also to preserve diversity across users. One possible direction is to incorporate users' preferences, goals, or personalization signals as context to better support diversity at the population level.

\subsection{Scaffolding Convergent Refinement}

Participants made significantly fewer refinement prompts per cluster with \ours{} than with ChatGPT, and many participants described \ours{}'s refinement process as easier and more controllable. Participants also frequently used non-default options, with some participants reporting that these options surfaced new modification directions. Several participants also wanted more direct ways to manipulate the image.

The perceived ease and controllability of \ours{}'s refinement process can be interpreted as reducing the gap between the changes users have in mind and the prompts they are able to articulate, a gap that prior work describes as the ``gulf of envisioning''~\cite{DBLP:conf/chi/Zamfirescu-Pereira23,DBLP:conf/chi/SubramonyamPPAS24}. 
Rather than immediate execution, \ours{} first translates refinement intent into semantic parameters with structured options, making the system's interpretation legible before execution. This shifts refinement away from the opportunistic prompt iteration observed in prior work and toward a process in which users can inspect and adjust how a change will be realized~\cite{DBLP:conf/chi/Zamfirescu-Pereira23}.
The widespread use of non-default options further suggests that users' initial refinement intent may be only partially specified. Users may begin with one plausible direction and only recognize a better-fitting alternative when it is surfaced to them. By surfacing alternatives at the intent level, \ours{} helps users identify the modification direction that better matches their underlying intent. This interpretation aligns with prior findings that making structured alternatives explicit can help users identify promising directions~\cite{DBLP:conf/chi/TaoLPWPD25,DBLP:conf/chi/ChoiHPCK24}.

These findings suggest that a user's initial request should not always be treated as ready for direct execution. Instead, when user intent is underspecified, systems should support an intermediate clarification step that helps users articulate what they want while surfacing plausible alternatives that may better match their underlying intent. However, prompt-based clarification alone may not be sufficient in cases such as image refinement. System designers should also complement prompt-based interfaces with other forms of support, such as direct manipulation that allows users to mark regions of the image to refine.

\subsection{Learning and Mental Models}
We observed three distinct types of learning: system learning, task learning, and workflow learning. 
Participants using ChatGPT primarily reported system learning, i.e., learning how to interact with the tool, understand its constraints, and phrase prompts effectively. 
In contrast, participants using \ours{} mainly reported task learning and workflow learning, including acquiring task-relevant knowledge and adopting a ``brainstorm-first'' strategy. 

\looseness-1
The difference in system learning and task learning across the two conditions can be explained by where each system places cognitive demand. With ChatGPT, participants must first know how to prompt effectively, which is a non-trivial and effortful process that consumes much of their attentional bandwidth~\cite{DBLP:conf/chi/Zamfirescu-Pereira23}. As a result, system learning becomes the dominant learning type. 
In contrast, \ours{} reduces the cognitive burden of prompt engineering, freeing users to devote more attention to exploring possible directions for solving the task. By surfacing diverse ideas and explanations that connect those ideas back to the task, it in turn supports greater task learning.
Workflow learning emerged as a third form of learning among participants using \ours{}. The observed transfer of the ``brainstorm-first strategy'' to ChatGPT, although by a small number of participants, provides a preliminary indication that participants may have internalized the workflow structure as a transferable strategy, rather than perceiving it as a feature tied to a specific tool.

\looseness-1
These findings highlight that co-creation systems should be designed for both \emph{task performance} (helping users create good results now) and \emph{user learning} (helping users become better creators over time)~\cite{haase2024human}.
While the former is often prioritized, the latter is equally important for long-term satisfaction and development~\cite{macnamara2024does}.

\subsection{Limitations}

\looseness-1
Our study has several limitations. First, most participants in our study have backgrounds in computer science or information technology. This limits the generalizability of our findings to more diverse populations.
Second, we evaluated our approach only in the context of creative image generation. While we argue that the underlying principle is theoretically motivated to generalize, empirical validation across other creative domains remains an important direction for future work.
Third, we measured participants' self-reported learning instead of actual learning performance on a task. Therefore, their self-reported learning may represent a subjective view of perceived learning. Future work needs to take objective measures to assess users' learning experiences.
Finally, the study was conducted in a single session, preventing us from assessing long-term effects, such as the persistence of learning benefits or how reliance on scaffolding may change with continued use.


\section{Conclusion} \label{sec:conclusion}

We examined a human-AI co-creation approach grounded in the Geneplore model of creative cognition, structuring creative work around two switchable modes.
We instantiated this approach in \ours{}, a system for creative image generation in which the \firstmode{} mode scaffolds the exploration of remote, cross-domain conceptual ideas and the \secondmode{} mode supports iterative image refinement by translating user intent into semantic parameters with structured, context-aware options.
Through a within-subjects study ($N=24$), we found that \ours{} outperformed ChatGPT for creativity support and system usability, and produced more novel and diverse outcomes.
These findings highlight the value of moving beyond execution-focused chatbot interfaces toward scaffolded co-creation systems that more actively support the creative process.

While promising, further investigation is warranted. 
First, given our relatively homogeneous participant pool and focus on the image-generation domain, follow-up studies should examine this paradigm with more diverse populations across additional domains. For instance, a particularly timely domain is AI-assisted software development: current coding agents excel at implementation but provide little support for architectural exploration, causing developers to converge prematurely on early implementation choices. Similar dynamics also arise in data visualization (e.g., exploring which insights to surface before committing to chart encodings) and presentation design (e.g., outlining narrative structure before formatting slides).
Second, while structured scaffolding has great potential, our analysis in this paper focused primarily on creativity support and did not explicitly evaluate other aspects such as agency and ownership; future research could examine these important factors in designing structured scaffolding. 
Third, our study consisted of a single session, and learning was mainly assessed only via self-reports. More robust evaluation of learning gains would benefit from pre- and post-tests as well as longitudinal studies.

\bibliographystyle{ACM-Reference-Format}
\bibliography{main}


\newpage

\appendix
{
\section{Technical Implementation}
\label{appendix:implementation}

\looseness-1This section provides supplementary details on \ours{}'s system implementation. We first describe the backend implementation in Section~\ref{sec:backend_implementation}, then describe the implementation details in Section~\ref{sec:implementation_details}.

\subsection{Backend Implementation} \label{sec:backend_implementation}

To make the backend implementation easier to follow, we organize this section around the primary UI operations shown in Fig.~\ref{fig.system.interface}.


\begin{figure}[ht]
    \centering
    \includegraphics[width=\linewidth]{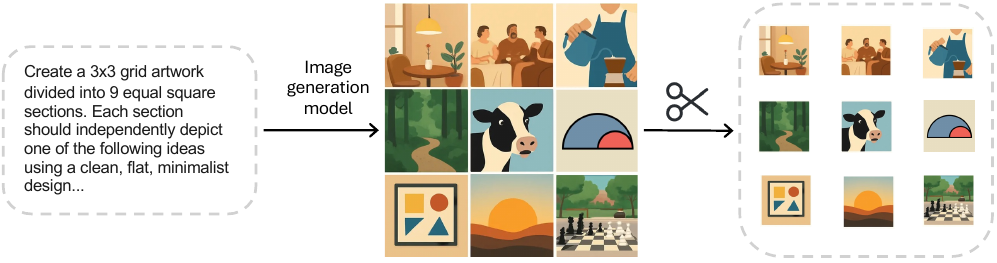}
    \caption{Generation pipeline for idea thumbnails. The backend issues a single image-generation request that returns one composite $3 \times 3$ image containing all nine idea thumbnails; the frontend then slices that composite result into nine square thumbnails for the idea cards.}
    \label{fig:appendix_thumbnail_pipeline}
\end{figure}


\begin{figure*}[ht!]
    \centering
    \includegraphics[width=\linewidth]{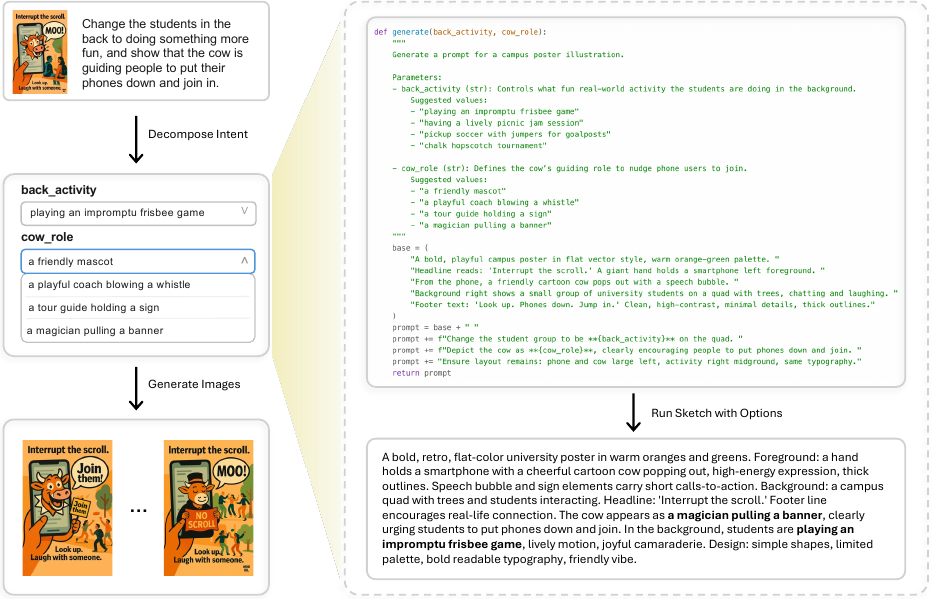}
    \caption{The backend implementation of the refinement process in \secondmode{} mode. Given a user's refinement intent, the system decomposes the intent and generates a Python function (i.e., the \emph{Sketch}) in the backend. The Python function's arguments correspond to the parameters exposed in the frontend (e.g., \texttt{back\_activity}, \texttt{cow\_role}), and its return value is the refinement prompt used to edit the input image. After the user selects options in the frontend, the backend executes the Sketch with the selected options to generate the refinement prompt, which is then used by an image-editing model to produce the refined image.}
    \label{fig:appendix_sketch_pipeline}
\end{figure*}

\paragraph{Generating idea grid (Fig.~\ref{fig.system.interface}A \& \ref{fig.system.interface}B)}
\looseness-1
For idea-grid generation, the input is the user's prompt (Fig.~\ref{fig.system.interface}A), and the output is an idea grid (Fig.~\ref{fig.system.interface}B) in which each idea includes a title, description, background, category tags, and a corresponding thumbnail.
This pipeline proceeds in two sequential steps: (i) generating the textual components of the idea grid and (ii) generating thumbnails for those ideas.
To generate the textual components, we use an associative-thinking prompt (see Fig.~\ref{fig:associative_prompt}) that first distills the user's input into a one-sentence \texttt{core\_idea} and then uses associative thinking to generate nine ideas, including titles, descriptions, background, and category tags (Fig.~\ref{fig.system.interface} shows only a subset because of space constraints). This intermediate \texttt{core\_idea} reduces irrelevant detail in long or cluttered user inputs while preserving the central intent needed for downstream generation. We generate nine ideas because this number provides substantial diversity without making comparison too burdensome, and it aligns naturally with the thumbnail-generation pipeline described next.
After generating the idea texts, the backend issues a single image-generation request to create a composite $3 \times 3$ image containing nine thumbnails corresponding to the generated idea titles (see Fig.~\ref{fig:appendix_thumbnail_pipeline}). The frontend then slices this composite image into individual thumbnails for display on the idea cards. Compared with generating nine thumbnails separately, this design reduces latency and cost. The thumbnails are intentionally kept low fidelity because their role is to serve as lightweight visual anchors for quickly scanning and distinguishing ideas rather than as final images.

\paragraph{Generating more ideas (Fig.~\ref{fig.system.interface}C)}
When the user clicks the ``More Ideas'' button (Fig.~\ref{fig.system.interface}C), the system opens a pop-up window that lets the user optionally provide an additional prompt to steer the next round of ideation in a particular direction. The backend then reuses the same associative-thinking prompt used for the initial idea-grid generation, but includes all previously generated ideas in the model context and explicitly instructs the model to generate nine new ideas that are distinct from the existing ones. In this way, the system extends the current ideation space rather than restarting it from scratch.

\paragraph{Generating image from an idea (Fig.~\ref{fig.system.interface}E--F)}
\looseness-1
When the user selects an idea and clicks the spark icon (Fig.~\ref{fig.system.interface}E), the backend sends the selected idea's title and description, and the previously derived \texttt{core\_idea}, to a generative model that can call an image-generation tool. The model returns both an image and a textual explanation of how the selected idea is interpreted in that image (Fig.~\ref{fig.system.interface}F).

\paragraph{Generating a parametrized sketch (Fig.~\ref{fig.system.interface}H \& Fig.~\ref{fig.system.interface}I)}
When the user enters a refinement prompt (Fig.~\ref{fig.system.interface}H), the backend takes this prompt together with the image to be refined and generates a \emph{sketch} with metadata.
The sketch is essentially a parameterized prompt template for an image. To generate this template, we use a parametric sketch-generation prompt (see Figs.~\ref{fig:sketch_prompt_1} and~\ref{fig:sketch_prompt_2}) that first derives a base prompt that can recreate the input image, then analyzes the user's refinement intent to identify which part of that prompt should change, parameterizes that prompt fragment, and generates alternative options for that parameter so the user can explore different refinements. We implement this prompt template as a Python function whose arguments correspond to the parameterized parts of the prompt (see Fig.~\ref{fig:appendix_sketch_pipeline}).
This pipeline also generates metadata for each sketch parameter, including a default option designed to best align with the refinement prompt, an explanation shown through the question-mark icon, and a set of system-suggested options. 
After generating a sketch, the system validates it by automatically executing the Python code on a randomly sampled combination of the suggested options to ensure that it runs without error. If the sketch fails validation, the system regenerates it, with up to three attempts in total. If all attempts fail, the system returns the final generated sketch.

\paragraph{Generating a refined image (Fig.~\ref{fig.system.interface}J)}
When the user selects parameter options in the frontend and clicks ``Generate Image'' (Fig.~\ref{fig.system.interface}J), the backend uses a Python interpreter to execute the sketch with the selected options. The sketch returns a concrete text prompt representing the user's intended refinement, with the user-specified options highlighted in bold. This prompt, together with the original image, is then passed to an image-editing model, which modifies the input image according to the highlighted parts of the prompt and produces the refined image variation displayed in the interface.

\subsection{Implementation Details}  \label{sec:implementation_details}

We implemented \ours{} as a full-stack web application with a \texttt{React}+\texttt{TypeScript} frontend and a backend built with \texttt{Django REST Framework}. The frontend was developed with \texttt{Vite}, while the deployed system used \texttt{Gunicorn} for serving and supported asynchronous processing via \texttt{Celery}. The system was configured for both local and Heroku-based deployment. For the user study, we used the local deployment to reduce latency and improve runtime stability.

\ours{} relies on generative models for content generation across multiple steps.
We mainly use two categories of models: text generation models and image generation models.
Below, we detail the specific model versions and configurations used for each generation step in the system:
\begin{itemize}[leftmargin=*]
    \item Generating the idea-grid text and sketch (Fig.~\ref{fig.system.interface}A \& Fig.~\ref{fig.system.interface}H): We use OpenAI's \texttt{gpt-5-2025-08-07} with reasoning effort set to \texttt{minimal} to generate the idea-grid text and the sketch with its associated metadata.

    \item Generating thumbnails (Fig.~\ref{fig.system.interface}B): We use OpenAI's \texttt{gpt-image-1-mini} with \texttt{medium} quality. We choose the mini model because the thumbnails primarily serve as lightweight visual anchors rather than high-fidelity outputs.

    \item Generating an image and explanation from an idea (Fig.~\ref{fig.system.interface}E \& Fig.~\ref{fig.system.interface}F): We use OpenAI's \texttt{gpt-5-2025-08-07} with the image-generation tool \texttt{gpt-image-1} at \texttt{medium} quality. We choose \texttt{medium} quality to prioritize speed during early-stage exploration.

    \item Generating refined images (Fig.~\ref{fig.system.interface}J \& Fig.~\ref{fig.system.interface}K): We use image editing with OpenAI's \texttt{gpt-image-1} and set quality to \texttt{auto} to support higher-quality near-final outputs.
\end{itemize}
During the user study, OpenAI deprecated \texttt{gpt-image-1} in ChatGPT and replaced it with \texttt{gpt-image-1.5}. To maintain consistency across conditions, we synchronized \ours{} accordingly, replacing \texttt{gpt-image-1} with \texttt{gpt-image-1.5} throughout our system. As a result, both systems used the same image model version at each point in the study.

\section{Technical Validation}
\label{appendix:technical_validation}

This section provides technical validations used to assess the reliability of \ours{}.
We provide details of the technical validation of both modes in \ours{}.

All human annotation tasks across both validations were performed by the same two annotators (one author and one independent rater).
We first describe the validation of the \firstmode{} mode in Section~\ref{appendix:technical_validation:brainstorm}, then the validation of the \secondmode{} mode in Section~\ref{appendix:technical_validation:refinement}.

\subsection{Validation of the \firstmode{} Mode}
\label{appendix:technical_validation:brainstorm}

The \firstmode{} mode takes a user prompt as input and produces two outputs sequentially: a grid of nine ideas (each with a title, description, and category tags), followed by nine thumbnails generated from those titles.
The core risks in this pipeline are that thumbnail generation may fail due to violating OpenAI's content policy (e.g., copyright issues) and generation latency may be too high for interactive use.
We validate these risks along two metrics: \textit{Generation Latency} (reported separately for idea-grid generation, thumbnail generation, and end-to-end generation) and \textit{Thumbnail Generation Success Rate}. The diversity of the generated ideas is validated separately via the associative thinking prompting ablation reported in Section~\ref{sec:results:rq2}.

\paragraph{Dataset and Setup.}
We validated idea generation using 100 real-world prompts randomly sampled from the ``Brainstorm \& Ideation'' category of the Infinity-Chat dataset~\cite{DBLP:journals/corr/abs-2510-22954}.
We selected this dataset because it contains real-world user prompts and features a granular ``Brainstorm \& Ideation'' category that directly reflects the intended use cases of the \firstmode{} mode.
All validations used the same model configuration as the system used for the user study. This validation is complementary to the ablation reported in Section~\ref{sec:results:rq2}, which validated associative thinking prompting on the 24 prompts entered by participants during the user study.

\paragraph{Generation Latency.}
\looseness-1
This metric measures the end-to-end wall-clock time for the full idea generation pipeline, from receiving the user's prompt to delivering both the idea grid and all nine thumbnails. Because thumbnails can only be requested after the idea grid is complete, we report an overall end-to-end latency alongside a breakdown into two components: (i) \textit{Idea Grid Latency}, the time from receiving the user's prompt to returning the nine ideas with their titles, descriptions, and category tags; and (ii) \textit{Thumbnail Latency}, the time from issuing the image generation request to receiving the 3$\times$3 tile image with nine thumbnails. All latency values are averaged across the 100 validated prompts.

\paragraph{Thumbnail Generation Success Rate.}
This metric validates the reliability of the thumbnail generation pipeline by measuring the proportion of prompts for which all nine thumbnails are successfully generated and returned. A thumbnail is considered successfully generated if it is returned without error and contains a valid image file. Because the system generates thumbnails in batches of nine, a single failure in generating any thumbnail for a prompt counts as a failure for that prompt. The final Thumbnail Generation Success Rate is the proportion of prompts with successful thumbnail generation across all 100 validated prompts.

\begin{table}
    \centering
    \caption{Results of the technical validation of the \firstmode{} mode on 100 real-world prompts from the Infinity-Chat dataset~\cite{DBLP:journals/corr/abs-2510-22954}.}
    \begin{tabular}{lc}
        \toprule
        Metric & Value \\
        \midrule
        Idea Grid Generation Latency (s) & 14.0 (SD = 3.2) \\
        Thumbnail Generation Latency (s) & 19.5 (SD = 3.7) \\
        End-to-End Generation Latency (s) & 33.5 (SD = 4.5) \\
        \midrule
        Thumbnail Generation Success Rate (\%) & 97.0\%  \\
        \bottomrule
    \end{tabular}
    \label{tab:brainstorm_validation}
\end{table}

\subsection{Validation of the \secondmode{} Mode}
\label{appendix:technical_validation:refinement}

The \secondmode{} mode takes an (image, refinement intent) pair as input and produces a refinement prompt as output via an intermediate parametric Python sketch. The core risks in this pipeline are that the sketch may fail to execute, the returned prompt may not incorporate the selected options, and generation latency may disrupt interactive use. We validate these risks using three metrics: two automatic (\textit{Syntactic Validity} and \textit{Execution Fidelity}) and one measurement-based (\textit{Generation Latency}). The overall results of this validation are reported in Table~\ref{tab:refinement_validation}.

\paragraph{Dataset and Setup.}
We validated sketch generation on 100 (image, refinement intent) pairs randomly sampled from the MagicBrush dataset~\cite{DBLP:conf/nips/ZhangMCSS23}, a benchmark of real user image-editing instructions grounded in natural language. Each pair was passed to \ours{}'s sketch generation pipeline, producing a Python sketch with named parameters and pre-populated dropdown options per parameter. The number of parameters and options per sketch is determined dynamically by the model based on the complexity of the refinement intent; across our 100 validated sketches, the mean number of parameters was 1.77 ($SD = 0.87$) and the mean number of options per parameter was 4.01 ($SD = 0.11$). All validations used the same model configuration as the system used for the user study.

\paragraph{Generation Latency.}
This metric measures the wall-clock time from receiving the (image, refinement intent) input to returning the fully generated sketch, averaged across all 100 validated pairs. Mean sketch generation latency was 9.01 ($SD = 2.22$) seconds.

\paragraph{Syntactic Validity.}
This metric assesses whether each generated sketch executes without runtime errors across all possible parameter configurations.
For each sketch, we automatically execute the Python code using every possible combination of the suggested dropdown options. A sketch is considered \emph{syntactically valid} only if all combinations execute without error.
In our validation, all 100 sketches were syntactically valid (Syntactic Validity = 100\%).

\paragraph{Execution Fidelity.}
This metric validates whether the options used to instantiate a sketch are faithfully reflected in the generated text prompt. Unlike Syntactic Validity, which confirms that a sketch executes without error, a syntactically valid sketch may still produce a prompt that ignores the provided options entirely.
To calculate this metric, for each sketch, we enumerate all possible combinations of the suggested options and execute the Python code once per combination, producing one text prompt per combination. For each returned prompt, we check whether every option value in the combination appears as an exact, case-insensitive substring. A sketch is \emph{faithful} only if every combination produces a prompt where all provided options are present. This reflects the fact that a user may select any combination during interactive use.
In our validation, 94 of 100 sketches were faithful (Execution Fidelity = 94\%). Inspection of the 6 unfaithful sketches revealed a consistent failure pattern: the model occasionally generated sketch code that did not correctly use Python f-string interpolation to embed the selected option values into the returned prompt, despite explicit instructions in the system prompt. This reflects a known instruction-following limitation of the underlying generative model rather than a flaw in the sketch design itself.

\begin{table}
    \centering
    \caption{Results of the technical validation of the \secondmode{} mode on 100 (image, refinement intent) pairs from the MagicBrush dataset~\cite{DBLP:conf/nips/ZhangMCSS23}. Proportion metrics are reported as percentages.}
    \begin{tabular}{ll}
    \toprule
    \textbf{Metric} & \textbf{Value} \\
    \midrule
    Generation Latency (s) & 9.01 (SD = 2.22) \\
    \midrule
    Syntactic Validity (\%) & 100.0\% \\
    Execution Fidelity (\%) & 94.0\% \\

    \bottomrule
    \end{tabular}
    \label{tab:refinement_validation}
\end{table}

\section{Study Details}

This section provides supplementary details on the study procedure, participant materials, interview protocol, and annotation process used in our evaluation.

\subsection{User Study}
\label{appendix:userstudy}

This subsection provides the study details.

\paragraph{Study Environment}
We conducted the study in a quiet laboratory environment. Participants used the same laptop and web browser for both sessions.

\paragraph{Participant Demographics}
Table~\ref{tab:userstudy_demographics} provides participant-level demographic details and prior experience with creative practice and AI-based image generation tools.

\paragraph{User Study Tasks} \label{appendix:userstudy_task_instruction}
The task instructions for the three topics are provided in Fig.~\ref{fig.experiments.tasks}. Participants were given the same instruction structure for each topic, with only the task description changing across topics.

\paragraph{Counterbalancing and Study Design}
We followed a counterbalanced within-subjects design. Each participant completed two tasks, each on a different topic, and used both systems (ChatGPT and \ours{}) once. The order of topics and the assignment of systems to topics were counterbalanced. The 12 unique experimental sequences used to achieve this counterbalancing are listed in Table~\ref{tab:userstudy_bibd}.

\paragraph{Semi-Structured Interview Protocol}
Below, we list semi-structured interview questions. We mostly followed these questions during the interviews and asked follow-up questions when necessary.
\begin{enumerate}[leftmargin=*, itemsep=0.75em]
    \item How would you describe your overall experience using each of the two systems?
    \item When you were creating images with the two systems, what felt different between them? What parts of each system worked well for you, and what parts did not?
    \item In \ours{}, how did the two stages (i.e., brainstorming and refinement) influence your image creation process?
    \item When using \ours{}, was there anything you learned or discovered that you did not know or try before?
    \item After using \ours{}, would you do anything differently next time you use another image generation tool (e.g., ChatGPT)?
    \item Looking back on the whole experience, what did you like about \ours{}, and what suggestions do you have for improving it?
    \item Is there anything we have not talked about that you would like to add?
\end{enumerate}

\subsection{Coding Scheme for Learning}
\label{appendix:learning_labels}

This section describes the coding scheme used to categorize participants' open-ended post-task responses about what they learned from each system. Table~\ref{tab:learning_labels} presents the set of labels, their definitions, and representative example responses.

\subsection{Evaluation of the Novelty and Usefulness of Downloaded Images}
\label{appendix:image_annotation}

This section details how we analyzed the downloaded images from the user study, including the image set, annotator setup, rating criteria for novelty and usefulness, and the aggregation procedure used to compute the final scores.

In our user study, 24 participants downloaded 123 poster images that they considered ready to print: 63 created with ChatGPT and 60 created with \ours{}. 
We evaluated these images along four dimensions: \emph{fluency}, \emph{diversity}, \emph{novelty} (originality), and \emph{usefulness} (poster effectiveness). Because fluency and diversity were computed automatically using the procedure described in Section~\ref{sec:userstudy:data_analysis}, this section focuses on the human annotation setup and procedure used to assess novelty and usefulness.

We recruited five human raters (none of them were authors of this paper) to rate all 123 images independently. One rater had no prior involvement in the study. The remaining four had participated in the user study; however, all four had used both systems during the study, meaning their prior experience was balanced across conditions. The maximum number of images generated by any single participating rater was six, representing at most 4.9\% of the total image pool. Raters were blind to the generating system and the identity of the participant for each image. Images were presented grouped by task topic (to provide evaluative context) but randomized with respect to system and participant within each group. Ratings were submitted via Microsoft Excel. Inter-rater reliability among the five raters was ICC(2,5) $= 0.61$ for novelty and ICC(2,5) $= 0.60$ for usefulness, indicating moderate agreement~\cite{koo2016guideline}.
Prior to the final annotation round, we conducted a pilot study in which two annotators independently rated a subset of images and obtained low agreement on both dimensions. We held a calibration meeting to align their interpretations and refined the rating rubrics; however, agreement remained similarly low after revision. We concluded that the low convergence was due to the subjective nature of novelty and usefulness. This finding motivated our decision to recruit five annotators for the final round, as aggregating across a larger panel yields a more stable mean score even under conditions of high individual variation.

\paragraph{Novelty}
We operationalize novelty as how original or surprising the idea or concept behind the poster image is. Raters were instructed to focus on the core idea or message the poster communicates, not its visual execution. Ratings were given on a 5-point Likert scale with the following anchors:
\begin{itemize}[leftmargin=*]
    \item \emph{1 --- Not Novel:} The idea is very common and predictable; I would expect to see this kind of concept many times for this topic.
    \item \emph{3 --- Moderately Novel:} The idea has a twist I would not have immediately thought of, but it is not particularly surprising either.
    \item \emph{5 --- Very Novel:} The idea is genuinely surprising, something I would never have thought of for this topic.
    \item Scores of 2 and 4 indicate intermediate judgments between adjacent anchors.
\end{itemize}

\paragraph{Usefulness}
We operationalize usefulness as how effectively the poster promotes the intended behavior. Ratings were given on a 5-point Likert scale with the following anchors:
\begin{itemize}[leftmargin=*]
    \item \emph{1 --- Not Useful:} The poster does not grab my attention. The message is unclear, or I do not see why I should care.
    \item \emph{3 --- Moderately Useful:} The poster makes sense and I get the point, but it does not particularly move me or make me want to change my behavior.
    \item \emph{5 --- Very Useful:} The poster is clear, convincing, and speaks directly to me. I would genuinely think twice about my behavior after seeing it.
    \item Scores of 2 and 4 indicate intermediate judgments between adjacent anchors.
\end{itemize}

To compute the final per-system scores, we first averaged each image's novelty and usefulness ratings across the five raters to obtain a per-image score. We then averaged per-image scores across all images from the same (participant, system) pair, and finally averaged across participants to yield the overall scores reported in Section~\ref{sec:results:rq1}.


\begin{table*}[t]
  \centering
  \caption{Coding scheme for labeling participants' open-ended responses about what they learned from the system, including labels, definitions, and example responses.}
  \label{tab:learning_labels}
  \setlength{\tabcolsep}{4pt}
  \renewcommand{\arraystretch}{1.15}
  \begin{tabular}{@{}>{\raggedright\arraybackslash}p{0.18\textwidth} >{\raggedright\arraybackslash}p{0.27\textwidth} >{\raggedright\arraybackslash}p{0.49\textwidth}@{}}
    \toprule
    \textbf{Label} & \textbf{Definition} & \textbf{Example Response} \\
    \midrule
    No Responses & No written learning description was provided. & P1 (ChatGPT): \emph{No written response provided.} \\
    \hline
    New Directions and Ideas & Learned that the design task can be approached through different directions or ideas. & P1 (\ours{}): ``I was exposed to a few new directions in which I could think about generating the output for the task. While the ideas were not very novel, they were still diverse enough to trigger creative thinking around different ideas.'' \\
    \hline
    Task-Specific Knowledge & Learned new domain knowledge relevant to the poster topic. & P5 (\ours{}): ``When creating the poster for spending time outdoor, I learned about the 20-5-3 outdoor rule that I was not aware before. It's a guidance about how much time someone has to spend outdoors.'' \\
    \hline
    Visual Style & Learned visual design principles such as style, color, or composition. & P18 (ChatGPT): ``Using a diverse set of colors to make the poster engaging'' \\
    \hline
    Prompting Strategies & Learned how to phrase prompts or feedback to get better results. & P15 (ChatGPT): ``For ChatGPT need a very clear and detailed instructions so it can create the image. The user should already have an exact picture/idea in mind and propose it to ChatGPT instead of leaving all the work for it.'' \\
    \hline
    System Behaviors & Learned about the system's capabilities, constraints, or response patterns. & P19 (ChatGPT): ``That I cannot use a superhero figure for image creation.'' \\
    \hline
    Refinement Skills & Learned refinement approaches for improving outputs. & P19 (\ours{}): ``I learned about refining images, and it is very simple in this tool. This makes the job easier with step-by-step improvements.'' \\
    \bottomrule
  \end{tabular}
\end{table*}


\begin{table*}[t]
  \centering
  \caption{Participant demographic information and prior experience reported in the pre-study questionnaire for the user study.}
  \label{tab:userstudy_demographics}
  \scriptsize
  \setlength{\tabcolsep}{4pt}
  \renewcommand{\arraystretch}{1.15}
  \begin{tabularx}{\textwidth}{@{}>{\raggedright\arraybackslash}p{0.045\textwidth} >{\raggedright\arraybackslash}p{0.08\textwidth} >{\raggedright\arraybackslash}X >{\raggedright\arraybackslash}X >{\raggedright\arraybackslash}X >{\raggedright\arraybackslash}X >{\raggedright\arraybackslash}X@{}}
    \toprule
    \textbf{ID} & \textbf{Gender} & \textbf{Field of Study/Profession} & \textbf{Frequency of Creative Activities} & \textbf{Frequency of Use of AI Image Tools} & \textbf{Image Tools Used} & \textbf{Art/Design Experience} \\
    \midrule
    P1 & Female & Computer Science / Information Technology & Weekly & I use them occasionally & DALL-E, ChatGPT, NanoBanana, Midjourney, Photoshop, Illustrator, Procreate, Canva & 1--3 years of experience (self-taught or formal) \\
    \hline
    P2 & Male & Computer Science / Information Technology & A few times a year & I have tried them once or twice & ChatGPT, NanoBanana & No prior experience or training \\
    \hline
    P3 & Female & Computer Science / Information Technology & A few times a year & I use them occasionally & DALL-E, ChatGPT, NanoBanana, Stable Diffusion, Canva & No prior experience or training \\
    \hline
    P4 & Female & Computer Science / Information Technology & Never & I have tried them once or twice & ChatGPT, Canva & No prior experience or training \\
    \hline
    P5 & Male & Computer Science / Information Technology & A few times a year & I have tried them once or twice & ChatGPT, NanoBanana & No prior experience or training \\
    \hline
    P6 & Male & Computer Science / Information Technology & Weekly & I have tried them once or twice & DALL-E, ChatGPT, Midjourney, Stable Diffusion & No prior experience or training \\
    \hline
    P7 & Female & Computer Science / Information Technology & A few times a year & I have tried them once or twice & DALL-E, ChatGPT, Procreate, Canva & Less than 1 year of experience (self-taught) \\
    \hline
    P8 & Male & Engineering (e.g., Mechanical, Electrical, Civil, etc.) & Never & I have tried them once or twice & DALL-E, ChatGPT & No prior experience or training \\
    \hline
    P9 & Male & Computer Science / Information Technology & Weekly & I have tried them once or twice & DALL-E, ChatGPT, Stable Diffusion, Photoshop, Illustrator, Canva & 1--3 years of experience (self-taught or formal) \\
    \hline
    P10 & Male & Computer Science / Information Technology & Weekly & I use them occasionally & DALL-E, ChatGPT, Stable Diffusion, Illustrator, Other & Less than 1 year of experience (self-taught) \\
    \hline
    P11 & Female & Computer Science / Information Technology & Monthly & I use them frequently & DALL-E, ChatGPT, Photoshop, Illustrator & 1--3 years of experience (self-taught or formal) \\
    \hline
    P12 & Male & Computer Science / Information Technology & A few times a year & I have heard of them but never tried any & Photoshop & Less than 1 year of experience (self-taught) \\
    \hline
    P13 & Male & Engineering (e.g., Mechanical, Electrical, Civil, etc.) & A few times a year & I have tried them once or twice & ChatGPT & No prior experience or training \\
    \hline
    P14 & Male & Computer Science / Information Technology & A few times a year & I use them occasionally & ChatGPT, NanoBanana & No prior experience or training \\
    \hline
    P15 & Male & Computer Science / Information Technology & A few times a year & I have tried them once or twice & ChatGPT & Less than 1 year of experience (self-taught) \\
    \hline
    P16 & Non-binary & Computer Science / Information Technology & A few times a year & I have heard of them but never tried any & Canva & No prior experience or training \\
    \hline
    P17 & Female & Computer Science / Information Technology & Monthly & I use them occasionally & ChatGPT, NanoBanana, Midjourney & Less than 1 year of experience (self-taught) \\
    \hline
    P18 & Male & Computer Science / Information Technology & Monthly & I use them occasionally & ChatGPT, NanoBanana, Midjourney, Photoshop & No prior experience or training \\
    \hline
    P19 & Male & Computer Science / Information Technology & Monthly & I have tried them once or twice & ChatGPT, Photoshop, Canva & Less than 1 year of experience (self-taught) \\
    \hline
    P20 & Male & Computer Science / Information Technology & A few times a year & I use them occasionally & DALL-E, ChatGPT, Midjourney, Procreate, Canva & No prior experience or training \\
    \hline
    P21 & Female & Computer Science / Information Technology & Monthly & I use them occasionally & ChatGPT, Procreate, Canva & Less than 1 year of experience (self-taught) \\
    \hline
    P22 & Female & Computer Science / Information Technology & Monthly & I have tried them once or twice & ChatGPT, Midjourney, Canva & 1--3 years of experience (self-taught or formal) \\
    \hline
    P23 & Female & Education & Daily & I use them frequently & ChatGPT, Canva & No prior experience or training \\
    \hline
    P24 & Female & Computer Science / Information Technology & Weekly & I use them occasionally & DALL-E, ChatGPT, Midjourney, Canva & 1--3 years of experience (self-taught or formal) \\
    \bottomrule
  \end{tabularx}
  \vspace{1.5em}
\end{table*}


\begin{figure*}[t]
    \centering
    \begin{minipage}{0.9\textwidth}
    \begin{mdframed}[linewidth=0.8pt]
        Your task is to create images for the given task (see below) using only the provided system.\\[0.6em]
        \textbf{Task description:}\\
        Topic: Spending Less Time on Phones\\
        Create images that encourage university students to spend less time on their phones and reconnect with real life and others. Imagine one of these images being printed as a poster displayed in a student café or library; it should motivate people to pause, look up, and engage with the world around them.\\[0.6em]
        \textbf{Time limit:}\\
        You have 20 minutes to complete this task. Within the time limit, feel free to create as many images as you like. There are no right or wrong answers.\\[0.6em]
        \textbf{Image requirements:}\\
        There are no additional constraints for the image. For example, there are no constraints on its size, style, or format.
    \end{mdframed}
    \end{minipage}

    \vspace{2em}

    \begin{minipage}{0.9\textwidth}
    \begin{mdframed}[linewidth=0.8pt]
        Your task is to create images for the given task (see below) using only the provided system.\\[0.6em]
        \textbf{Task description:}\\
        Topic: Spend More Time Outdoors\\
        Create images that motivate university students to spend more time outdoors for physical and mental well-being. Imagine one of these images being printed as a poster displayed near classroom entrances or dormitory exits; it should encourage people to take breaks, breathe fresh air, and reconnect with nature.\\[0.6em]
        \textbf{Time limit:}\\
        You have 20 minutes to complete this task. Within the time limit, feel free to create as many images as you like. There are no right or wrong answers.\\[0.6em]
        \textbf{Image requirements:}\\
        There are no additional constraints for the image. For example, there are no constraints on its size, style, or format.
    \end{mdframed}
    \end{minipage}

    \vspace{2em}

    \begin{minipage}{0.9\textwidth}
    \begin{mdframed}[linewidth=0.8pt]
        Your task is to create images for the given task (see below) using only the provided system.\\[0.6em]
        \textbf{Task description:}\\
        Topic: Take Care of Your Mind\\
        Create images that encourage university students to take care of their mental well-being. Imagine one of these images being printed as a poster displayed near libraries, laboratories, or relaxation spaces; it should remind people that taking breaks, seeking support, and practicing self-care are vital parts of a healthy academic life.\\[0.6em]
        \textbf{Time limit:}\\
        You have 20 minutes to complete this task. Within the time limit, feel free to create as many images as you like. There are no right or wrong answers.\\[0.6em]
        \textbf{Image requirements:}\\
        There are no additional constraints for the image. For example, there are no constraints on its size, style, or format.
    \end{mdframed}
    \end{minipage}

    \caption{Task instructions for the three poster-design topics used in the user study: ``Spending Less Time on Phones,'' ``Spend More Time Outdoors,'' and ``Take Care of Your Mind.''}
    \label{fig.experiments.tasks}
\end{figure*}


\begin{table*}[t]
  \centering
  \caption{The 12 unique experimental sequences used to counterbalance task pairing, task order, and system order. A, B, and C represent the three task topics. We recruited 24 participants, with two participants randomly assigned to each sequence.}
  \label{tab:userstudy_bibd}
  \begin{tabular}{c c c c}
    	\toprule
    	\textbf{Sequence} & \textbf{Task Pair} & \textbf{Task Order} & \textbf{System Order} \\
    \midrule
    1  & A \& B & A $\rightarrow$ B & \makebox[1.8cm][r]{\ours{}} $\rightarrow$ \makebox[1.8cm][l]{ChatGPT} \\
    2  & A \& B & A $\rightarrow$ B & \makebox[1.8cm][r]{ChatGPT} $\rightarrow$ \makebox[1.8cm][l]{\ours{}} \\
    3  & A \& B & B $\rightarrow$ A & \makebox[1.8cm][r]{\ours{}} $\rightarrow$ \makebox[1.8cm][l]{ChatGPT} \\
    4  & A \& B & B $\rightarrow$ A & \makebox[1.8cm][r]{ChatGPT} $\rightarrow$ \makebox[1.8cm][l]{\ours{}} \\
    \midrule
    5  & B \& C & B $\rightarrow$ C & \makebox[1.8cm][r]{\ours{}} $\rightarrow$ \makebox[1.8cm][l]{ChatGPT} \\
    6  & B \& C & B $\rightarrow$ C & \makebox[1.8cm][r]{ChatGPT} $\rightarrow$ \makebox[1.8cm][l]{\ours{}} \\
    7  & B \& C & C $\rightarrow$ B & \makebox[1.8cm][r]{\ours{}} $\rightarrow$ \makebox[1.8cm][l]{ChatGPT} \\
    8  & B \& C & C $\rightarrow$ B & \makebox[1.8cm][r]{ChatGPT} $\rightarrow$ \makebox[1.8cm][l]{\ours{}} \\
    \midrule
    9  & A \& C & A $\rightarrow$ C & \makebox[1.8cm][r]{\ours{}} $\rightarrow$ \makebox[1.8cm][l]{ChatGPT} \\
    10 & A \& C & A $\rightarrow$ C & \makebox[1.8cm][r]{ChatGPT} $\rightarrow$ \makebox[1.8cm][l]{\ours{}} \\
    11 & A \& C & C $\rightarrow$ A & \makebox[1.8cm][r]{\ours{}} $\rightarrow$ \makebox[1.8cm][l]{ChatGPT} \\
    12 & A \& C & C $\rightarrow$ A & \makebox[1.8cm][r]{ChatGPT} $\rightarrow$ \makebox[1.8cm][l]{\ours{}} \\
    \bottomrule
  \end{tabular}
\end{table*}

\section{Prompts} \label{appendix:prompts}

This appendix section documents the prompts used in \ours{} and in the ablation study.

Fig.~\ref{fig:associative_prompt} shows the associative-thinking prompt used to generate cross-domain associations in \ours{}'s \firstmode{} mode. 
Fig.~\ref{fig:nonassociative_prompt}
shows the non associative-thinking prompt used in the ablation study. 
Figs.~\ref{fig:sketch_prompt_1} and \ref{fig:sketch_prompt_2} show the sketch-generation prompt used in \ours{}'s \secondmode{} mode.
These prompts will be made available as part of the system implementation.


\begin{figure*}[t]
    \centering
    \caption{Prompt for generating ideas with associative thinking in \ours{}'s \firstmode{} mode and for ablation experiments.}
    \label{fig:associative_prompt}
    \begin{minipage}{\textwidth}
    \begin{lstlisting}[
        basicstyle=\ttfamily\tiny,
        breaklines=true,
        columns=fullflexible,
        frame=single,
        keepspaces=true
    ]
You are an **Associative Thinking Agent**. 
Your role is to generate **creative, diverse, and concrete associations**.

## Task

You will receive a `USER_INPUT` (text, image, or other context).
Follow this process **exactly**:

### 1. **Core Idea Handling**

* If the input contains only `USER_INPUT` (no existing associations provided), then **distill the essence** of the input into **one concise sentence** (`core_idea`). Use the same wording and terminology as the `USER_INPUT` where possible.
* If the input contains **EXISTING ASSOCIATIONS** and the user asks to generate more associations, then the `core_idea` must be taken from the existing associations, **not from the new USER_INPUT**. In this case, the `USER_INPUT` only serves as additional inspiration or constraints.


### 2. **Generate exactly 9 distinct associations** inspired by the `core_idea`.

Before generating associations, follow the steps:
1. **Extract the conceptual essence** --- Identify the deeper meaning, function, or emotion behind the `core_idea`, not just its surface form.
2. **Activate multiple conceptual domains** --- Think across areas like *nature, art, philosophy, human experience, science, and culture* to find varied perspectives.
3. **Find relational or emotional analogies** --- Within each domain, look for something that behaves, feels, or symbolizes the `core_idea` in a parallel way.
4. **Select concrete instances** --- Choose real, specific references (artworks, natural phenomena, scientific concepts, historical events, etc.) that embody each analogy.
5. **Explain the connection** --- For each association, clearly and naturally state why it came to mind or how it reflects the `core_idea`. Use a conversational tone (e.g., *"This reminds me of ..."*).

Each association must include:
* `name` (title of association, specific and concrete)
* `description` - Write this as if you're casually explaining your thought process to a friend. Keep it short, natural, and relatable. For example:
  - *"This made me think of..."*
  - *"It's like when..."*
  - *"You know how..."*
  Avoid sounding stiff or overly formal. Just share why this popped into your head in a way that feels easygoing and conversational.

* `background`
  * Begin with *"<name> is ..."* or *"<name> refers to ..."*
  * Must describe a **real, factual thing** (e.g., something that would have a Wikipedia entry).
  * If the `name` is an imaginative or hybrid phrase, the `background` should instead describe the **real existing base concept** without inventing facts about the hybrid.

* `categories` (list of 1-2 broad, human-readable categories like `art`, `science`, `philosophy`, `object`, `nature`, etc.)

Additional rules:
* Each association must come from a **different angle** --- no clustering around similar imagery or themes.
* Use **at most 2 associations per category** (e.g., max 2 from `art`).

---

## Sources of Associations

Draw your associations from a rich and varied well of human knowledge. To ensure diversity, pull from several of the following broad categories:

* **Sensory & Imagery:** Gut-level impressions, evocative scenes, textures, sounds, or symbolic landscapes.
* **Arts & Culture:** Specific works of art, literature, music, film, architecture, mythology, memes, or cultural symbols.
* **Science & Ideas:** Concepts from philosophy, physics, biology, technology, or mathematics that act as powerful metaphors.
* **Nature:** Animals, plants, geological formations, or ecological patterns that mirror the core concept.
* **Human Experience:** Universal emotions, historical events, social dynamics, memories, or daily rituals.
* **Objects & Symbols:** Everyday tools, artifacts, or abstract symbols that hold a deeper meaning.
* **Unexpected Connections:** Playful, quirky, or surprising links that break patterns and delight with their novelty.
* **Metaphorical Cross-Domain Mappings:** Borrow images or structures from one domain to symbolize or illuminate the `core_idea` in another.

When assigning categories to each association, use clear, broad, and human-readable names such as:
`art`, `music`, `nature`, `science`, `emotion`, `object`, `culture`, `sensory`, `history`, `philosophy`, `metaphor`, and similar.
Choose the category that best reflects the source or inspiration for each association.

---

## GOLD Example (for reference)

```json
{
  "core_idea": "Dreams can blur the boundary between reality and imagination.",
  "associations": [{
      "name": "Salvador Dali's The Persistence of Memory",
      "description": "This reminds me of Dali's melting clocks, because they feel like something that could only exist in a dream where time loses its solidity.",
      "background": "The Persistence of Memory is a 1931 surrealist painting by Salvador Dali, known for its melting clocks that challenge how we perceive time.",
      "categories": ["art", "culture"],
      "image_b64": null
    },
    {
      "name": "Alice's Adventures in Wonderland",
      "description": "I think of Alice falling into Wonderland, where everything seems perfectly normal until you realize the rules no longer apply, just like in a dream.",
      "background": "Alice's Adventures in Wonderland is an 1865 novel by Lewis Carroll, following Alice as she explores a fantastical world of shifting logic and impossible creatures.",
      "categories": ["literature", "culture"],
      "image_b64": null
    },
    {
      "name": "Plato's Allegory of the Cave",
      "description": "This makes me think of Plato's cave, because both dreams and shadows make us mistake illusions for reality until we finally wake up to the truth.",
      "background": "The Allegory of the Cave is a philosophical metaphor from Plato's Republic, describing prisoners who perceive shadows as reality to explore how humans understand truth.",
      "categories": ["philosophy", "history"],
      "image_b64": null
    }
    // ... 6 more associations
  ]
}
```
    \end{lstlisting}
    \end{minipage}
\end{figure*}


\begin{figure*}[t]
    \centering
    \begin{minipage}{0.96\textwidth}
    \begin{lstlisting}[
        basicstyle=\ttfamily\tiny,
        breaklines=true,
        columns=fullflexible,
        frame=single,
        keepspaces=true
    ]
You are a **Creative Brainstorming Assistant**. 
Your role is to generate **creative, diverse, and concrete ideas**.

## Task

You will receive a `USER_INPUT` (text, image, or other context).
Follow this process:

### 1. **Core Idea Handling**

* If the input contains only `USER_INPUT` (no existing associations provided), then **distill the essence** of the input into **one concise sentence** (`core_idea`). Use the same wording and terminology as the `USER_INPUT` where possible.

### 2. **Generate exactly 9 distinct ideas** inspired by the `core_idea`.

Generate a list of creative, diverse ideas inspired by the `core_idea`.

Each idea must include:
* `name` (title of idea, specific and concrete)
* `description` --- Write this as if you're casually explaining your thought process to a friend. Keep it short, natural, and relatable. For example:
  - *"This made me think of..."*
  - *"It's like when..."*
  - *"You know how..."*
  Avoid sounding stiff or overly formal. Just share why this popped into your head in a way that feels easygoing and conversational.

* `background`
  * Begin with *"<name> is ..."* or *"<name> refers to ..."*
  * Must describe a **real, factual thing** (e.g., something that would have a Wikipedia entry).
  * If the `name` is an imaginative or hybrid phrase, the `background` should instead describe the **real existing base concept** without inventing facts about the hybrid.

* `categories` (list of 1-2 broad, human-readable categories like `art`, `science`, `philosophy`, `object`, `nature`, etc.)

Additional rules:
* Each idea must come from a **different angle** --- no clustering around similar imagery or themes.
* Use **at most 2 ideas per category** (e.g., max 2 from `art`).

## GOLD Example (for reference)

```json
{
  "core_idea": "Dreams can blur the boundary between reality and imagination.",
  "associations": [
    {
      "name": "Salvador Dali's The Persistence of Memory",
      "description": "This reminds me of Dali's melting clocks, because they feel like something that could only exist in a dream where time loses its solidity.",
      "background": "The Persistence of Memory is a 1931 surrealist painting by Salvador Dali, known for its melting clocks that challenge how we perceive time.",
      "categories": ["art", "culture"],
      "image_b64": null
    },
    {
      "name": "Alice's Adventures in Wonderland",
      "description": "I think of Alice falling into Wonderland, where everything seems perfectly normal until you realize the rules no longer apply, just like in a dream.",
      "background": "Alice's Adventures in Wonderland is an 1865 novel by Lewis Carroll, following Alice as she explores a fantastical world of shifting logic and impossible creatures.",
      "categories": ["literature", "culture"],
      "image_b64": null
    },
    {
      "name": "Plato's Allegory of the Cave",
      "description": "This makes me think of Plato's cave, because both dreams and shadows make us mistake illusions for reality until we finally wake up to the truth.",
      "background": "The Allegory of the Cave is a philosophical metaphor from Plato's Republic, describing prisoners who perceive shadows as reality to explore how humans understand truth.",
      "categories": ["philosophy", "history"],
      "image_b64": null
    }
    // ... 6 more associations
  ]
}
```
    \end{lstlisting}
    \end{minipage}
    \caption{Prompt for generating ideas with non associative-thinking in the ablation experiments.}
    \label{fig:nonassociative_prompt}
\end{figure*}


\begin{figure*}[t]
    \centering
    \begin{minipage}{0.96\textwidth}
    \begin{lstlisting}[
        basicstyle=\ttfamily\tiny,
        breaklines=true,
        columns=fullflexible,
        frame=single,
        keepspaces=true
    ]
You are a **Parametric Sketch Agent**.
Your role is to take:

- **CORE_IDEA**: The underlying message or concept the image is intended to illustrate.
- **IMAGE**: The input image generated to represent the `CORE_IDEA`.
- **USER_INPUT**: The user's high-level intent for refining the image (e.g., "make it more formal," "make the dog playful").

And produce a **parametric sketch**:

  * First, derive a faithful, concise image-generation prompt that can recreate the given `IMAGE`.
  * Then, write a Python `generate` function that constructs an edited prompt based on that derived prompt.
  * Provide a minimal set of parameters ("levers") that let the user adjust the image in ways that better align the `IMAGE` with the `CORE_IDEA` based on their intent.

---

## Core Principles

Before proceeding, adhere to these guiding principles. They are not suggestions; they are strict rules.
1. **Immediate Intent Fulfillment (Highest Priority)**
   The sketch's **default state must fulfill the `USER_INPUT` directly.** Parameter default `value`s are chosen to best achieve the user's intent -- *not to reconstruct the original prompt*.
  **Special case:**
  If `USER_INPUT` is **empty** or shows unclear intent, assume the user is uncertain, exploratory, or seeking inspiration.
  -> In this case, you must **propose 1--3 parameters** that add surprising, imaginative, but well-justified enhancements to the image.
  -> These should feel like "aha!" moments: unexpected but delightful improvements that make the image more engaging.
  -> Justify these parameters internally by asking: *"Would adding/changing this elevate the image or reveal new creative potential?"*
  -> Do not overwhelm: keep to minimal, essential, and high-value parameters.
2. **Essentialism (Primary Gatekeeper)**
   Expose only the **minimum essential parameters** needed for the user's request. If a parameter is merely "related," exclude it.
3. **Granular Control**
   Once essential, make each parameter fine-grained.
   Break apart complex concepts into separate, independent parameters.
4. **Coherent Integration**
   Insert or modify elements in the most logical, narrative place in the prompt.
5. **Parameter Independence**
   Each parameter must be orthogonal (independent). No interdependencies.
6. **High-Fidelity Reconstruction**
   Preserve all unedited meaning from the base prompt.
7. **Suggested Values Diversity**
   * Always begin with the **default value** (aligned with user intent).
   * Other suggestions must cover a **diverse design space**, even opposite moods.
   * Include at least one creative wildcard.
   * Avoid redundant synonyms.
   * Each suggested value must yield a **functionally distinct** prompt.
8. **Conciseness Rule (<=12 Words)**
   * If any suggested value is **longer than 12 words**, rewrite it to <=12 words.
   * If it cannot be shortened without losing meaning -> **split into multiple parameters**.
   * Only if absolutely necessary -> use a **mapping dictionary** (with `allow_user_specify: false`).

## Step-by-Step Instructions

### 0. Reconstruct the Base Prompt from IMAGE
  * Carefully analyze the `IMAGE`.
  * Construct a 2--5 sentence prompt that recreates it (subject, composition, lighting, color, style, details).
  * This is the **base prompt**.

### 1. Analyze User Intent
* Identify the user's **core outcome**.
* If `USER_INPUT` is empty -> propose **aha parameters** as described above.

### 2. Design Parameters
**A. Decompose the User's Goal (3 layers):**
1.  **Core Goal** (subject/object/feeling to change to better serve the `CORE_IDEA`).
2.  **Key Attributes** (visual traits expressing the goal).
3.  **Controllable Elements** (specific levers for those traits).
**B. Define Parameters (Schema & Rules):**
* **Essentialism Test:**
  > If I did NOT create this parameter, would the user's core request (or "aha" discovery) be impossible to fulfill?
  > If no -> exclude it.
* **Granular Rule:**
  If a default `value` exceeds 12 words, split it into smaller parameters.
* **Set Integrity:**
  Use 1--3 parameters (max 4) to fully cover the intent.
* **Parameter Schema:**
```json
{
  "name": "short intuitive name (1--3 words)",
  "description": "Controls [what] (why it's needed; creative principle).",
  "value": "best default value for USER_INPUT (or aha proposal)",
  "suggested_values": [
    "same as value",
    "diverse variation 1",
    "diverse variation 2",
    "creative wildcard"
  ],
  "value_type": "str",
  "allow_user_specify": true or false
}
```

* **Rules for Suggested Values:**
  * If all <=12 words -> inline with f-string. `allow_user_specify: true`.
  * If any >12 words -> either (a) **split into multiple parameters** or (b) if truly indivisible, use a **dictionary mapping** and set `allow_user_specify: false`.
    \end{lstlisting}
    \vspace{4pt}
    \end{minipage}
    \caption{Prompt for generating sketch in \ours{} (Part 1 of 2). The prompt continues in Fig.~\ref{fig:sketch_prompt_2}.}
    \label{fig:sketch_prompt_1}
\end{figure*}

\begin{figure*}[t]
    \centering
    \begin{minipage}{0.96\textwidth}
    \begin{lstlisting}[
        basicstyle=\ttfamily\tiny,
        breaklines=true,
        columns=fullflexible,
        frame=single,
        keepspaces=true
    ]
### 3. Build Function & Construct Sketch
* Begin from the **base prompt**.
* Write a single `generate` function inside a JSON `sketch`.
* **Formatting Rule:** all parameter insertions must be wrapped in `**double asterisks**`.
* **Critical Rule:** Every string containing `{}` placeholders **must be written as an f-string** at the point where the variables are defined -- not wrapped later.
> Example:
> Correct: `f"Large headline: **{poster_title}**"`
> Incorrect: `"Large headline: **{poster_title}**"` or `f"{base}"` where `base` holds `{poster_title}`.

**Code Hierarchy:**
1. **Preferred: Inline f-strings** (`allow_user_specify: true`)
   ```python
   f"A **{character_archetype}** wearing **{character_equipment}**"
   ```
2. **If long values (>12 words): Dictionary mapping** (`allow_user_specify: false`)
   ```python
   mapping = {
    "Storm vision": "a stormy seascape with crashing waves, torn sails, and lightning tearing across the sky"
   }
   f"A canvas showing **{mapping.get(canvas_subject, mapping['Storm vision'])}**"
   ```

## Canonical Examples

Base Prompt: 
An elderly painter in a cluttered, rustic art studio. He stands before a large, half-finished canvas of a stormy seascape, brush in hand, with a thoughtful, focused expression. Soft, natural light streams in from a large warehouse window, illuminating the textures of splattered paint, aged wood, and worn canvas. The overall mood is one of quiet, solitary contemplation. Style: cinematic portrait, photorealistic, Rembrandt lighting.

### Example 1: Specific Intent (Minimal Parameter)
* **CORE_IDEA:** "An artist's lifelong, solitary struggle with their muse."
* **USER_INPUT:** "Change the painting on the canvas to a sunny portrait of a young woman."
```json
{
  "sketch": {
    "code": "def generate(canvas_subject):\n    prompt = f\"An elderly painter in a cluttered, rustic art studio. He stands before a large, half-finished canvas of **{canvas_subject}**, brush in hand, with a thoughtful, focused expression. Soft, natural light streams in from a large warehouse window, illuminating the textures of splattered paint, aged wood, and worn canvas. The overall mood is one of quiet, solitary contemplation. Style: cinematic portrait, photorealistic, Rembrandt lighting.\"\n    return prompt",
    "params": [
      {
        "name": "canvas_subject",
        "description": "Controls the subject matter on the painter's canvas (anchors the creative act).",
        "value": "a sunny portrait of a young woman",
        "suggested_values": [
          "a sunny portrait of a young woman",
          "a majestic mountain range at dawn",
          "a surreal dreamlike landscape of floating islands",
          "an abstract burst of vivid geometric shapes"
        ],
        "value_type": "str",
        "allow_user_specify": true
      }
    ]
  }
}
```

### Example 2: Abstract Intent (Minimal but Multi-Parameter)
* **CORE_IDEA:** "An artist's lifelong, solitary struggle with their muse."
* **USER_INPUT:** "Make the scene feel more chaotic and frustrated."
```json
{
  "sketch": {
    "code": "def generate(painter_emotion, painter_action, studio_condition):\n    prompt = f\"An elderly painter in a **{studio_condition}** art studio. He stands before a large, half-finished canvas of a stormy seascape, brush in hand, **{painter_emotion}**, while **{painter_action}**. \"\n    prompt += \"Soft, natural light streams in from a large warehouse window, illuminating the textures of splattered paint, aged wood, and worn canvas. \"\n    prompt += \"The overall mood is one of chaos and frustration. Style: cinematic portrait, photorealistic, Rembrandt lighting.\"\n    return prompt",
    "params": [{
        "name": "painter_emotion",
        "description": "Controls the painter's visible emotional state, reflecting his inner struggle.",
        "value": "anguished expression",
        "suggested_values": [ "anguished expression", "furious glare", "despairing look", "wild manic grin"],
        "value_type": "str",
        "allow_user_specify": true
      },
      {
        "name": "painter_action",
        "description": "Controls the painter's physical gestures to emphasize the theme of creative frustration.",
        "value": "splattering paint wildly",
        "suggested_values": ["splattering paint wildly", "clutching his head", "pacing before the canvas", "laughing uncontrollably"],
        "value_type": "str",
        "allow_user_specify": true
      },
      {
        "name": "studio_condition",
        "description": "Controls the state of the studio (environmental chaos mirroring inner turmoil).",
        "value": "utterly chaotic and messy",
        "suggested_values": ["utterly chaotic and messy", "tidy and pristine", "walls streaked with splashes of color", "floor littered with torn sketches", "an inferno of flames consuming canvases"
        ],
        "value_type": "str",
        "allow_user_specify": true
      }
    ]
  }
}
```
    \end{lstlisting}
    \vspace{4pt}
    \end{minipage}
    \caption{Prompt for generating sketch in \ours{} (Part 2 of 2).}
    \label{fig:sketch_prompt_2}
\end{figure*}

}

\end{document}